\documentclass[prb]{revtex4}
\usepackage{amssymb}

\usepackage{graphicx}

\begin{document}

\author{Alejandro Cabo*}
\affiliation{ Facultad de F\'{\i}sica, Pontificia Universidad
Cat\'olica de Chile, Vicu\~na Mackenna 4860, 6904411 Macul,
Santiago, Chile} \affiliation{Abdus Salam International Center for
Theoretical Physics, Strada Costiera, 11-34014, Trieste , Italy}
\author{Francisco Claro}
\affiliation{ Facultad de F\'{\i}sica, Pontificia Universidad
Cat\'olica de Chile, Vicu\~na Mackenna 4860, 6904411 Macul,
Santiago, Chile }

{\flushright
 Preprint ICTP, IC/2003/60(2003)}
\bigskip
\bigskip

\title{Analytic mean-field Hall crystal solution at $\nu =1/3$: composite fermion
like sub-bands and correlation effects}
\date{May 5, 2001}

\begin{abstract}
An analytic solution of the Hartree-Fock problem for a 2DEG at
filling 1/3 and half an electron per unit cell is presented. The
Coulomb interaction dynamically breaks the first Landau level in
three narrow sub-bands, one of which is fully occupied and the
other empty, as in the composite fermion model. The localized
orbitals associated to the Bloch like single electron
wavefunctions are nearly static, resembling the angular momentum
eigenstates within a Landau level for non-interacting fermions.
Strong correlations are expected owing to the large charge density
overlap between neighboring plaquettes. A numerical evaluation
brings the cohesive energy close to that of the best present day
models. It is also found that correlations are long range,
requiring over 50 particles spread over a finite sample to
approach convergence. Since presently allowed exact calculations
are far from this number, the question of how relevant the
considered wave-function is for the description of the ground
state of the 2DEG system remains open.
\medskip

\noindent PACS numbers: 73.43.Cd, 73.43.-f

\ \bigskip

* On leave of absence from {\it Theoretical Physics Group, Instituto de
Cibern\'etica, Matematica y F\'{i}sica, Calle E, No. 309, Vedado,
La Habana, Cuba. }
\end{abstract}

\maketitle

\section{Introduction}

It is well accepted today that the fractional quantum Hall effect results
from the formation of a strongly correlated spin polarized electron liquid
in the lowest Landau level, that crystallizes below filling 1/7 \cite
{stormer,laughlin,book}. It is intriguing, however, that a mean field charge
density wave solution exists at all fillings, called a Hall crystal (HC),
which yields in a natural way the odd-denominator rule that characterizes
the effect.\cite{clar1,clar2} The state is characterized by a unit cell that
contains a fractional charge of even denominator, such as e/2. The detailed
investigation of this class of mean field states was stopped largely due to
their higher energy than that of the liquid, and the very small correlation
correction obtained by Yoshioka and Lee (YL) for the Wigner solid (WS),
which accomodates one whole electron per plaquette.\cite{yosh} However,
while in the latter case there is little overlap of electrons from
neighboring cells, in the former the charge density forms continuous ridges
between cells opening the way for an appreciable increase of the cohesive
correlation energy.\cite{cabo1} This idea was first explored in the simplest
case of $1/2$ filling.\cite{cabclaper} The results confirmed the effect
yielding a second order correlation correction an order of magnitude larger
than for the best WS state. The same line of thinking was also recently
considered by Mikhailov.(\cite{mikha}).

In this work we report results for the correlation energy of the HC at $1/3$
filling. The work rests on an analytic solution of the mean field problem
constructde using symmetry considerations. We show that the
electron-electron interaction breaks the single particle degeneracy of the
lowest Landau level (LLL), splitting it into three remarkably flat bands.
All bands contain the same number of states so that only one is filled while
the other two are empty, separated by a large gap. While the HC unit cell
traps 3/2 flux quanta of the original field, it is only pierced by one flux
quanta of the reduced field $B*=B_{1/3}-B_{1/2},$ where $B_{1/3}$ and $%
B_{1/2}$ are the fields at $1/3$ and $1/2$ filling factor, respectively. The
narrowness of the occupied band and this latter fact suggest that the
magnetic Wannier states are approximate solutions of the problem and behave
as nearly free quasiparticles filling the LLL of the effective field $B*,$
much as the composite fermion model predicts.\cite{book,clar3}

We compute the correlation energy using the YL method, that is, evaluating
the second order perturbation correction to the Hartree-Fock (HF) solution.
Two methods are used, one employing a Monte Carlo approach for the space
integrals over a large sample, and the other by computing in a discrete
momentum space using a set defined by periodic boundary conditions applied
to a comparatively small sample. As for the half filling case, both methods
yield an energy correction of about 10\%, an order of magnitude larger than
that for the WS state\cite{yosh}, confirming the earlier conjecture that at
all fillings the high electron overlap of the HC enhances significantly the
cohesive energy corrections\cite{cabo1,mikha}, thus re-opening the case for
this state as the proper precursor in a first principles perturbative
approach to the true gound state of the system in the thermodynamic limit.\

One further outcome of our calculations is that over 50 particles
would be required in order to approach convergence, way above
current permissible sizes used in first principles numerical
evaluations normally performed with up to 12 particles,  number
that has been extended to treat at most 27 particles \cite{yosh1}.
This may be understood again looking at the large inter cell
overlap through hexagonal ridges, which seems to point out the
importance of cooperative ring exchange effects in the
system.\cite{arov} To be operative, these need large sample sizes
and would be missed by such numerical diagonalizations of the
hamiltonian. If the true ground state retains the essential
features of its HF approximation, then one would expect the Hall
crystal to be the proper mean field precursor due to its higher
correlation energy than other mean field
solutions.\cite{tesanovic}

In Sec. 2 the single particle Hartree-Fock hamiltonian is diagonalized
analytically. Section 3 is devoted to the second order correction os the
mean field solution. Finally, in Section 4 we present our conclusions.
Details of the calculation are given in appendices A-D. In particular, in
Appendix B a formula is derived expressing the single particle Hartree-Fock
Hamiltonian in terms of the magnetic translations associated to an arbitrary
periodicity lattice of the\ HF\ problem. This formula, up to our knowledge,
is absent in the literature.

\section{Bloch reduction of the HF problem at $\nu =\frac{1}{3}$}

\subsection{One particle Hamiltonian and magnetic translations}

We consider a 2DEG in a strong perpendicular magnetic field. We are
interested in an analytic treatment of the Hartree-Fock problem at filling
1/3. We start out by writing the Hartree-Fock hamiltonian in the compact form

\begin{equation}
H_{HF=}\sum_{Q}v(\mathbf{Q})\exp (-\frac{r_{o}^{2}\mathbf{Q}^{2}}{4}%
)T_{r_{o}^{2}\ \mathbf{n\times Q}},  \label{hamiltonian}
\end{equation}
where the coefficients $v(\mathbf{Q})$ are given by
\begin{equation}
v(\mathbf{Q})=2\pi r_{o}^{2}\rho (\mathbf{Q})\exp (\frac{r_{o}^{2}\mathbf{Q}%
^{2}}{4})\left( \frac{1-\delta _{\mathbf{Q,0}}}{r_{o}\mid \mathbf{Q\mid }}%
\exp (-\frac{r_{o}^{2}\mathbf{Q}^{2}}{4})-\sqrt{\frac{\pi }{2}}I_{o}(\frac{%
r_{o}^{2}\mathbf{Q}^{2}}{4})\right) \frac{e^{2}}{\varepsilon _{o}r_{o}},
\label{poten}
\end{equation}
where $\varepsilon _{o}$ is the background dielectric constant. The magnetic
translation operators T are defined in Appendix A, while form (1) is derived
in Appendix B. The particle density in real space $\rho (\mathbf{x})$ is
assumed to be periodic under displacements covering the triangular lattice
defined by the vectors
\begin{eqnarray}
\mathbf{R} &=&r_{1}\mathbf{a}_{1}+r_{2}\mathbf{a}_{2},\ \ r_{1},r_{2}=0,\pm
1,\pm 2,...;  \label{perlatt} \\
\mathbf{a}_{1} &=&\sqrt{\frac{6\pi }{\sqrt{3}}}(1,0)\ r_{o},  \nonumber \\
\mathbf{a}_{2} &=&\sqrt{\frac{6\pi }{\sqrt{3}}}(\frac{1}{2},\frac{\sqrt{3}}{2%
})\ r_{o}.  \nonumber
\end{eqnarray}
The Fourier components of the density$\ $are defined as
\begin{equation}
\rho (\mathbf{Q})=\frac{1}{A_{cell}}\int \mathbf{dx\ }\rho (\mathbf{x})\exp
(i\mathbf{Q.x}),  \label{chardenfour}
\end{equation}
where $A_{cell}$ is the unit cell area
\begin{equation}
A_{cell}=\mathbf{n.a}_{1}\mathbf{\times a}_{2}=3\pi r_{o}^{2}.
\end{equation}
Here $\mathbf{n}$ is a unit vector normal to the electron gas plane. Through
this cell traverse a flux $\frac{3}{2}$ in units of the magnetic flux
quantum $\phi _{o}=hc/e$. Associated with the above real space lattice is
the reciprocal lattice
\begin{eqnarray}
\mathbf{Q} &=&Q_{1}\mathbf{s}_{1}+Q_{2}\ \mathbf{s}_{2} \\
Q_{1},Q_{2} &=&0,\pm 1,\pm 2,...  \nonumber \\
\mathbf{s}_{1} &=&-\frac{2}{3r_{o}^{2}}\mathbf{n\times a}_{2},  \nonumber \\
\mathbf{s}_{2} &=&\ \ \frac{2}{3r_{o}^{2}}\mathbf{n\times a}_{1},  \nonumber
\\
\mathbf{s}_{i}.\mathbf{a}_{j} &=&2\pi \ \delta _{ij}.  \nonumber
\end{eqnarray}

\subsection{Block diagonalization of the HF hamiltonian}

We next show that it is possible to find a basis in which the matrix
representation of (1) has a diagonal form composed of simple 3-dimensional
blocks. In addition, the functions have such a structure that they
automatically furnish the translation symmetry of the total density over the
lattice (2). For this purpose we consider the basis functions $\varphi _{%
\mathbf{p}}(\mathbf{x})\ $defined in Appendix A, constructed over the
lattice with primitive vectors
\begin{eqnarray}
\mathbf{b}_{1} &=&\mathbf{a}_{1},  \label{bas32} \\
\mathbf{b}_{2} &=&\frac{2}{3}\mathbf{a}_{2}.  \nonumber
\end{eqnarray}
The magnetic translations then have the form
\begin{equation}
T_{r_{o}^{2}\ \mathbf{n\times Q}}=T_{-\frac{2}{3}Q_{2}\mathbf{b}_{1}+Q_{1}%
\mathbf{b}_{2}}.  \label{trans}
\end{equation}
Since the flux piercing the unit cell of the lattice (2) is not an integral
number of flux quanta, the set of translation operators obtained by varying
the integers $Q_{1}$and $Q_{2}$ in (\ref{trans}) do not commute and it is
not possible to find common eigenfunctions to all of them. The basis we
shall construct decomposes in a set of three dimensional subspaces, closing
each of them under the action of translations for all values of $\mathbf{Q}$%
. \

A first step in finding the basis is to define a set of eigenfunctions of a
translation in the vector $\mathbf{a}_{2},$ which is a period of the
density. For a given value of the momentum $\mathbf{p}$ such orbitals may be
written as
\begin{equation}
\chi _{\mathbf{p}}^{\sigma }(\mathbf{x})=\frac{1}{\sqrt{2}}\left( \varphi _{%
\mathbf{p}}(\mathbf{x})+\frac{\sigma }{\exp (-i\mathbf{a}_{2}\mathbf{.p})}T_{%
\mathbf{a}_{2}}\varphi _{\mathbf{p}}(\mathbf{x})\right) ,\sigma =\pm 1.
\end{equation}
Using the formulas in Appendix A it can be readily proven that these
functions satisfy the eigenvalue relation
\begin{equation}
T_{\mathbf{a}_{2}}\chi _{\mathbf{p}}^{\sigma }(x)=\sigma \exp (-i\mathbf{a}%
_{2}\mathbf{.p})\ \chi _{\mathbf{p}}^{\sigma }(\mathbf{x}).
\end{equation}
The two values of $\sigma $ appearing in these equations will play an
important role in what follows. They will allow us to impose the periodicity
of the density under the shifts in $\mathbf{a}_{1}$ and $\mathbf{a}_{2},$ in
spite of the impossibility of obtaining a basis of eigenvectors of all the
magnetic translations in the lattice. It should be stressed that the range
of values of $\mathbf{p}$ defining independent functions in the new basis
have been reduced in half, the two values of $\sigma $ compensating for this
reduction. The restriction comes from the singular property of the basis $%
\left\{ \varphi _{\mathbf{p}}\right\} ,$ that a magnetic translation is
fully equivalent to a shifting of the momenta argument as implied by the
relation
\begin{equation}
T_{\mathbf{R}}\varphi _{\mathbf{p}}(\mathbf{x})=\mathcal{F}_{\mathbf{p}}(%
\mathbf{R})\varphi _{\mathbf{p}+\frac{2e}{\hbar c}\mathbf{A(R)}}(\mathbf{x}),
\end{equation}
where $\mathcal{F}_{\mathbf{p}}(\mathbf{R})$ is a pure phase factor. See
Appendix A and Ref (16) for the justification of this relation. It directly
implies that the shift done in $\mathbf{a}_{2}$ in constructing the new
basis precisely changes the momentum $\mathbf{p}$ of the particle in $-%
\mathbf{s}_{1}/2$. Therefore, what has been basically done is to form linear
combinations of the original functions associated with different values of
the momentum. In the new basis the magnetic translation in $\mathbf{a}_{1}$
has the simple effect
\begin{equation}
T_{\mathbf{a}_{1}}\chi _{\mathbf{p}}^{\sigma }(\mathbf{x})=\exp (-i\mathbf{a}%
_{1}\mathbf{.p})\ \chi _{\mathbf{p}}^{-\sigma }(\mathbf{x}),
\end{equation}
that is, it merely changes the sign of $\sigma $ and multiplies it by a
phase factor.

As the next step let us employ the fact that, although the functions are not
eigenvectors of translations in $\mathbf{a}_{1}$, they are eigenfunctions of
the double sized translations in $2\mathbf{a}_{1}$. This is because its
effect, when considered as two consecutive shifts in $\mathbf{a}_{1},$ have
the simple result of making two consecutive changes of sign of $\sigma $
that reproduce the original function. Therefore, if for $\mathbf{p}$ and $%
\sigma $ fixed we construct the triplet of states formed by the function $%
\chi _{\mathbf{p}}^{\sigma }(\mathbf{x})$ and the other two obtained by a
pair of successive translations in the vector -$2\mathbf{a}_{1}/3$, the
operation of performing a translation in an arbitrary multiple of these
vectors leaves the triplets invariant.

A specific linear combination within each triplet which is also an
eigenfunction of the translation in -$2\mathbf{a}_{1}/3$ can be obtained by
constructing the new basis
\begin{equation}
\chi _{\mathbf{p}}^{(r,\sigma )}(\mathbf{x})=\sum_{s=-1,0,1}c_{r}^{s}(%
\mathbf{p})T_{-\frac{2}{3}s\,\mathbf{b}_{1}}\ \chi _{\mathbf{p}}^{\sigma }(%
\mathbf{x}),\ r=-1,0,1.  \label{functions}
\end{equation}
After solving the linear equations for the constants $c_{r}^{s}$ obtained by
imposing the condition that these functions be solutions of the eigenvalue
problem
\[
T_{-\frac{2}{3}\,\mathbf{b}_{1}}\chi _{\mathbf{p}}^{(r,\sigma )}(x)=\lambda
\,\chi _{\mathbf{p}}^{(r,\sigma )}(\mathbf{x})
\]
one finds
\begin{eqnarray}
\lambda ^{(r)}(\mathbf{p}) &=&\exp (\frac{2}{3}i\,\mathbf{p.a}_{1}+\frac{%
2\pi r}{3}) \\
c_{s}^{r}(\mathbf{p}) &=&\frac{1}{\sqrt{3}}\exp (-\frac{2}{3}i\,\mathbf{p.a}%
_{1}s-\frac{2\pi irs}{3})\,  \nonumber \\
r,s &=&-1,0,1.  \nonumber
\end{eqnarray}
Substituting in (\ref{functions}) yields
\begin{eqnarray}
\chi _{\mathbf{p}}^{(r,\sigma )}(\mathbf{x}) &=&\frac{1}{\sqrt{3}}%
\sum_{s=-1,0,1}\exp (-\frac{2}{3}i\,\mathbf{p.a}_{1}s-\frac{2\pi irs}{3})T_{-%
\frac{2}{3}s\,\mathbf{b}_{1}}\ \chi _{\mathbf{p}}^{\sigma }(\mathbf{x}), \\
r &=&-1,0,1  \nonumber \\
\sigma  &=&\pm 1  \nonumber \\
\mathbf{p} &\equiv &\mathbf{p}+n\frac{\mathbf{s}_{1}}{2}+m\frac{\mathbf{s}%
_{2}}{2},\,\,n,m=0,\pm 1,\pm 2,...  \label{brillouin}
\end{eqnarray}
The last relation expresses that, modulo a phase factor, the states of the
new basis are equivalent upon a shift of their momenta $\mathbf{p}$ in any
linear combination with integer coefficients of half the unit cell vectors
of the reciprocal lattice corresponding to the periodicity of the density.
The periodicity of the states under the shifts in $\mathbf{s}_{1}/2$ was
discussed above, and the one related with $\mathbf{s}_{2}/2$ similarly
follows from the relation (\ref{phase}) in Appendix A, expressing the
equivalence of a magnetic translation with a shift in momentum. The
functions just defined have an alternative and more compact form given by
\begin{eqnarray}
\chi _{\mathbf{p}}^{(r,\sigma )}(x) &=&\frac{1}{\sqrt{6}N_{\mathbf{p}%
}^{(3,2)}}\sum_{\mathbf{m}}\exp (i\mathbf{P}^{(\mathbf{p},r,\sigma )}.%
\mathbf{m}+\frac{5\pi i}{6}m_{1}m_{2})T_{\mathbf{m}}\phi (\mathbf{x})\text{
,\ \ }  \label{altern} \\
N_{\mathbf{k}}^{(3,2)} &=&\sqrt{N_{\phi _{0}}}\sqrt{\sum_{\mathbf{\ell }%
}\left( -1\right) ^{\ell _{1}\ell _{2}}\exp (i\ \mathbf{k}.\mathbf{\ell -}%
\frac{\mathbf{\ell }^{2}}{4r_{0}^{2}}\mathbf{)\ }}, \\
\mathbf{\ell } &=&\ell _{1}(3\mathbf{c}_{1}\text{)}+\ell _{2}\text{\ (2}%
\mathbf{c}_{2}),
\end{eqnarray}
where the effective momenta and the new elementary lattice of vectors $%
\mathbf{m}$ over which the sum is performed are given by
\begin{eqnarray}
\mathbf{P}^{(\mathbf{p},r,\sigma )} &=&\mathbf{p}-r\ \mathbf{s}_{1}+\frac{%
\sigma -1}{2}\frac{3}{2}\mathbf{s}_{2}, \\
\mathbf{m} &=&m_{1}\mathbf{c}_{1}+m_{2}\mathbf{c}_{2},  \nonumber \\
\mathbf{c}_{1} &=&\frac{\mathbf{a}_{1}}{3},  \nonumber \\
\mathbf{c}_{2} &=&\frac{\mathbf{a}_{2}}{3}.  \nonumber
\end{eqnarray}
The double sum (\ref{altern}) can be evaluated to obtain an explicit formula
in terms of the Elliptic Theta functions as follows
\begin{eqnarray}
\chi _{\mathbf{p}}^{(r,\sigma )}(\mathbf{x}) &=&\frac{\exp (\mathbf{-}\frac{%
\mathbf{x}^{2}}{4r_{0}^{2}}\mathbf{)}}{\sqrt{6}\sqrt{2\pi r_{0}^{2}}N_{%
\mathbf{p}}^{(3,2)}}\times   \nonumber \\
&&\sum_{\beta =0,...5}\sum_{\alpha =0,1}\exp {\LARGE (}27\pi i\tau
_{1}\alpha ^{2}+\pi i\tau _{1}\beta ^{2}+a_{2}\beta +2\pi i\alpha (\frac{%
3a_{2}}{\pi i}-\frac{3a_{1}}{2\pi i}-\frac{15\alpha }{2}+(\frac{9\tau _{1}}{2%
}-\frac{5}{4})\beta ){\LARGE )}\times   \nonumber \\
&&\text{ }\Theta _{3}(\frac{6a_{2}}{\pi i}-\frac{3a_{1}}{\pi i}-(15-54\tau
_{1})\alpha +2(\frac{9\tau _{1}}{2}-\frac{5}{4})\beta \mid 108\text{ }\tau
_{1})\times \Theta _{3}(\frac{a_{1}}{2\pi i}+\frac{5\alpha }{2}+(\frac{\tau
_{1}}{2}+\frac{5}{12})\beta )\mid \tau _{1}),
\end{eqnarray}
where
\begin{eqnarray*}
a_{1} &=&i\mathbf{P}^{(\mathbf{p},r,\sigma )}.\mathbf{c}_{1}+\frac{1}{%
2r_{o}^{2}}(\mathbf{c}_{2-i}\ \mathbf{n\times c}_{1}).\mathbf{x} \\
a_{2} &=&i\mathbf{P}^{(\mathbf{p},r,\sigma )}.\mathbf{c}_{2}+\frac{1}{%
2r_{o}^{2}}(\mathbf{c}_{2-i}\ \mathbf{n\times c}_{2}).\mathbf{x} \\
&&\tau _{1=}\frac{i}{6\sqrt{6}}.
\end{eqnarray*}

Let us inspect now the action of a magnetic translation by $\frac{2}{3}%
\mathbf{a}_{2}$ on the new functions. If such a transformation leaves the
triplets invariant, then the matrix reduction of the Hartree-Fock
Hamiltonian will follow. One has,\

\begin{equation}
T_{\frac{2}{3}\mathbf{a}_{2}}\chi _{\mathbf{p}}^{(r,\sigma )}(\mathbf{x})=%
\frac{1}{\sqrt{3}}\sum_{s=-1,0,1}\exp (-\frac{2}{3}i\mathbf{\,p.a}_{1}s-%
\frac{2\pi irs}{3})T_{\frac{2}{3}\mathbf{a}_{2}}T_{-\frac{2}{3}s\,\mathbf{b}%
_{1}}\ \chi _{\mathbf{p}}^{\sigma }(\mathbf{x}).
\end{equation}
But after using (\ref{commutation}) for changing the order of the two
operators within the sum, it follows that
\begin{equation}
T_{\frac{2}{3}a_{2}}\chi _{\mathbf{p}}^{(r,\sigma )}(\mathbf{x})=\lambda _{%
\mathbf{p}}(\frac{2}{3}\mathbf{a}_{2})\,\chi _{\mathbf{p}}^{([r-1],\sigma )}(%
\mathbf{x}),
\end{equation}
where the square bracket defines the number among the set \{-1,0,1\} that is
equivalent, modulo 3, to the integer in the argument.

For fixed $\mathbf{p}$ and $\sigma $ the matrix elements of the Hamiltonian (%
\ref{hamiltonian}) in the new basis can readily be found to have the form
\begin{equation}
h_{\mathbf{p},\sigma }^{(r^{\prime },r)}=\sum_{Q}v(\mathbf{Q})\exp (-\frac{%
\mathbf{Q}^{2}r_{o}^{2}}{4})\exp \left( i\,\mathbf{p.n\times Q}\,\,r_{o}^{2}+%
\frac{2\pi i}{3}Q_{2}(r+Q_{1})\right) \delta _{r^{\prime },[r-Q_{1}]}.
\label{heigen}
\end{equation}
The problem has thus been reduced to the self-consistent diagonalization of
a three dimensional matrix for each value of momenta $\mathbf{p}$ and
parameter $\sigma $. \ The basis can be checked to have the following set of
transformations properties,
\begin{eqnarray}
T_{2\mathbf{c}_{1}}\chi _{\mathbf{p}}^{(r,\sigma )}(\mathbf{x}) &=&\ \exp
(-2i\ \mathbf{p.c}_{1}-\frac{2\pi r\ i}{3}\,)\chi _{\mathbf{p}}^{(r,\sigma
)}(\mathbf{x}),  \label{symmetry} \\
T_{2\mathbf{c}_{2}}\chi _{\mathbf{p}}^{(r,\sigma )}(\mathbf{x}) &=&\ \exp
(-2i\ \mathbf{p.c}_{2}\,)\chi _{\mathbf{p}}^{([r-1],\sigma )}(\mathbf{x}),
\nonumber \\
T_{3\mathbf{c}_{1}}\chi _{\mathbf{p}}^{(r,\sigma )}(\mathbf{x}) &=&\ \exp
(-3i\ \mathbf{p.c}_{1}\,)\chi _{\mathbf{p}}^{(r,-\sigma )}(\mathbf{x}),
\nonumber \\
T_{3\mathbf{c}_{2}}\chi _{\mathbf{p}}^{(r,\sigma )}(\mathbf{x}) &=&\ \sigma
\exp (-3i\ \mathbf{p.c}_{2}\,)\chi _{\mathbf{p}}^{(r,\sigma )}(\mathbf{x}),
\nonumber \\
I\ \chi _{\mathbf{p}}^{(r,\sigma )}(\mathbf{x}) &=&\chi _{-\mathbf{p}%
}^{(-r,\sigma )}(\mathbf{x}),  \nonumber
\end{eqnarray}
where the parity transformation $I$ is defined as usual by $I\ \chi _{%
\mathbf{p}}^{(r,\sigma )}(\mathbf{x})=\chi _{\mathbf{p}}^{(r,\sigma )}(-%
\mathbf{x}).$ \ As shown in Appendix A, from these symmetry properties it
follows that the density associated with any Slater determinant constructed
by selecting one orbital within each triplet has exact periodicity under
shifts in all vectors $\mathbf{R}$.

In order to find the solution of the mean field problem by an iterative
technique it is sufficient to make an ansatz for the density in the first
step, and then diagonalize numerically the matrices for a sufficiently high
partition of the reduced Brillouin cell momenta $\mathbf{p}$ defined by (\ref
{brillouin}). By selecting the normalized lowest energy state within each
three dimensional quantum mechanical problem, the Fourier components of the
density corresponding to the new step should be constructed. Following its
definition, it can be done by means of the following expression,
\begin{eqnarray}
\rho (\mathbf{Q}) &=&\frac{1}{A_{cell}}\sum_{p,\sigma =\pm 1}\int \mathbf{dx|%
}\sum_{r}g_{r}^{0}(\mathbf{p})\chi _{\mathbf{p}}^{(r,\sigma )}(\mathbf{x}%
)|^{2}\exp (i\mathbf{Q.x})  \label{density} \\
&=&\frac{\,\exp (-\frac{\mathbf{Q}^{2}r_{o}^{2}}{4})}{A}\sum_{\mathbf{p}%
}\sum_{\sigma =\pm 1}\exp (i\,\mathbf{p.n\times Q}\,\,r_{o}^{2}+\frac{2\pi i%
}{3}Q_{2}Q_{1})\times   \nonumber \\
&&\sum_{r,r^{\prime }=-1,01}g_{r^{\prime }}^{0}(\mathbf{p}))^{*}g_{r}^{0}(%
\mathbf{p})\,\exp (-\frac{2\pi i\,r}{3}Q_{2})\delta _{r^{\prime },[r+Q_{1}]}%
{\LARGE .}  \nonumber
\end{eqnarray}
This formula can be obtained by evaluating the Gaussian integrals appearing
after substituting the expansions defining the functions $\chi _{\mathbf{p}%
}^{(r,\sigma )}.\ $The coefficients $g_{r}^{0}$, $g_{r}^{1}$ and $g_{r}^{2}$%
, define the components of the eigenvectors of the single particle HF
Hamiltonian in the basis of states $\chi $. They fix the wave-functions of
the filled band and the empty bands as
\begin{eqnarray}
\Phi _{\mathbf{p}}^{(b,\sigma )}(\mathbf{x}) &=&\sum_{r=-1,0,1}g_{r}^{b}(%
\mathbf{p})\chi _{\mathbf{p}}^{(r,\sigma )}(\mathbf{x})  \label{solution} \\
b &=&0,1,2.  \nonumber
\end{eqnarray}
Here $b=0$ labels the filled band in each triplet and $b=1,2$ label the two
empty bands.

\subsection{ Functions that vanish at the origin are exact solutions}

In order to proceed within an analytical context let us consider the
observation from former numerical studies, that the particle density when
the number of flux quanta piercing the unit cell is a half integer
essentially vanishes at all lattice points.\cite{clar2} Then, let us first
assume that the density rigorously vanishes at this set of points. If so is
the case, the wave-function of any of the filled states should then also
vanish at those points. This requirement follows from the fact that the
Hartree-Fock particle density is a sum over the individual densities of all
occupied orbitals,

\begin{equation}
\sum_{\mathbf{p},\sigma =\pm 1}{\LARGE |}\sum_{r}g_{r}^{0}(\mathbf{p})\chi _{%
\mathbf{p}}^{(r,\sigma )}(\mathbf{x}){\LARGE |}^{{\LARGE 2}}=\rho (\mathbf{x}%
).  \label{charge}
\end{equation}
We can then use this property to fix the coefficients of the wave functions
within each triplet.

After imposing the vanishing conditions at the origin, the coefficients
defining the functions (\ref{solution}) are fully determined and take the
form
\begin{eqnarray}
g_{0}^{0}(\mathbf{p}) &=&\frac{1}{\mathcal{N}^{*}_{\mathbf{p}}}
\nonumber \\
g_{-1}^{0}(\mathbf{p}) &=&-\frac{1}{\mathcal{N}^{*}_{\mathbf{p}}}\frac{%
\chi _{\mathbf{p}}^{(0,+)}(0)\ \chi _{\mathbf{p}}^{(+1,-1)}(0)-\chi _{%
\mathbf{p}}^{(0,-1)}(0)\ \chi _{\mathbf{p}}^{(-1,-1)}(0)}{\chi _{\mathbf{p}%
}^{(-1,+1)}(0)\ \chi _{\mathbf{p}}^{(+,-1)}(0)-\chi _{\mathbf{p}%
}^{(+1,+1)}(0)\ \chi _{\mathbf{p}}^{(-1,-1)}(0)} \\
g_{+1}^{0}(\mathbf{p}) &=&-\frac{1}{\mathcal{N}^{*}_{\mathbf{p}}}\frac{%
\chi _{\mathbf{p}}^{(0,-1)}(0)\ \chi _{\mathbf{p}}^{(-1,+1)}(0)-\chi _{%
\mathbf{p}}^{(0,+1)}(0)\ \chi _{\mathbf{p}}^{(-1,-1)}(0)}{\chi _{\mathbf{p}%
}^{(-1,+1)}(0)\ \chi _{\mathbf{p}}^{(+1,-1)}(0)-\chi _{\mathbf{p}%
}^{(+1,+1)}(0)\ \chi _{\mathbf{p}}^{(-1,-1)}(0)}  \nonumber \\
1 &=&|g_{0}^{0}(\mathbf{p})|^{2}+|g_{-1}^{0}(\mathbf{p})|^{2}+|g_{1}^{0}(%
\mathbf{p})|^{2}.  \nonumber
\end{eqnarray}
Note that the coefficients $g$ are all independent of $\sigma $. This
completes the definition of our functions. That they are true solutions of
the Hartree-Fock problem is proven in Appendix C.

\begin{figure}[tbp]
\includegraphics[width=4in]{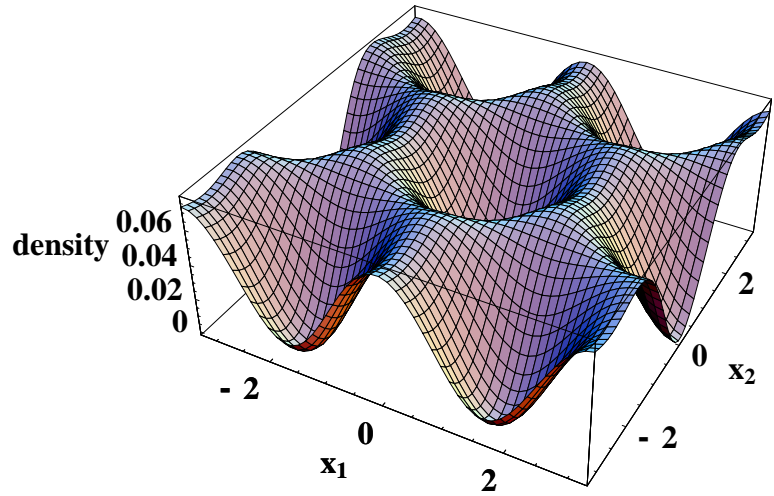}
\caption{The density of particles associated with the HC state. Its
structure indicates the high overlapping of the localized Wannier-like
states in terms of which the HF determinant always can be expressed. }
\label{fig:densi}
\end{figure}

The particle density may now be computed replacing these functions in Eq. (%
\ref{charge}). The real space particle density thus obtained is shown in Fig.%
\ref{fig:densi}.

A main property to be noticed in this figure is the formation of sharp
hexagonal channels surrounding the low density regions at the center of
which the vanishing density occurs. These structures mark the difference
with the Wigner solid whose unit cell encloses one flux quanta. The charge
density in this latter instance is made up essentially of well localized
gaussians centered at each lattice point. In our case there is strong
overlap, suggesting that cooperative ring exchange involving many unit cells
is a large contribution to the correlation energy.

\begin{figure}[tbp]
\includegraphics[width=3.5in]{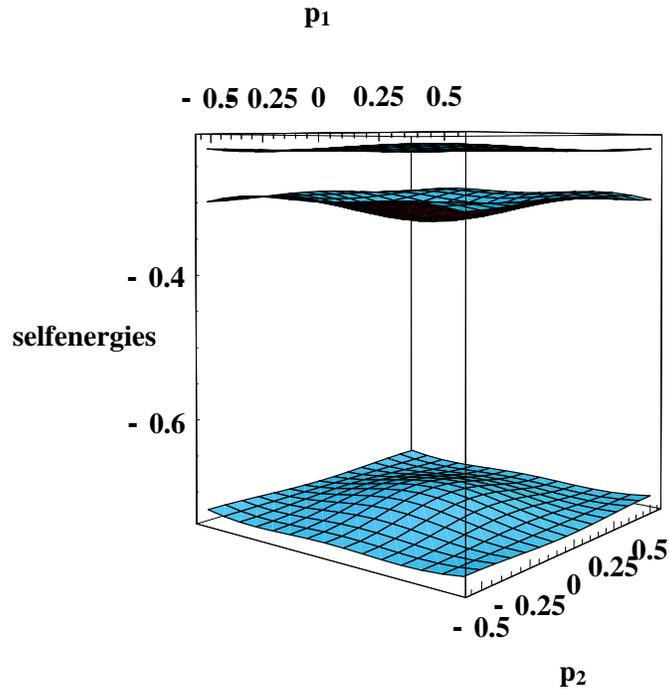}
\caption{Triplet of narrow bands in which the Landau level is splitted by
the action of the Coulomb interaction. }
\label{fig:bands}
\end{figure}

Further, the insertion of the calculated density in the eigenvalue equation
associated with the matrix representation of the Hamiltonian in each
triplet, Eqs. (\ref{poten}) and (\ref{heigen}), allows for the evaluation of
the one particle spectrum of the system. As it was expected, three energy
bands appear, each associated with a value of the index r and covering the
full range of $\mathbf{p}$ within each triplet. We also note that states
associated with $\sigma =\pm 1$ turn out to be degenerate. The bands
dispersion relations are illustrated in Fig.~\ref{fig:bands}. 

Note the narrowness of the bands as compared with the gaps separating them.
This fact leads to the idea that in this mean field approximation the
Coulomb interaction, although breaking the degeneracy of the first Landau
level, it reorganizes the states in three equally populated separate sets
that again are aproximately degenerate, as if they were Landau levels of a
renormalized problem. Since at 1/3 filling one band is full and the other
two empty, one expects the electrons to behave dynamically similarly to
filling one, except for a different effective mass, as the composite fermion
model suggests.\cite{cabclaper,clar3} The magnetic Wannier states are
expected to be approximate solutions, playing the role of the angular
momentum states in the non-interacting problem at filling one.

These properties seem to corroborate the possibility of tracing a link
between weakly interacting composite fermions and the Bloch or Wannier
orbitals in the mean field solutions considered earlier by one of us.\cite
{clar1,clar2} To finish this section we note that the mean field energy per
particle obtained from the above formalism for the solution we have
constructed confirms the value reported in an earlier numerical computation,%
\cite{clar2}
\begin{equation}
{\Large \epsilon }=-0.362\frac{e^{2}}{\varepsilon _{o}r_{o}}.  \label{energy}
\end{equation}


\section{Correlation energy in second order}

In order to obtain the energy correction to second order of perturbation
theory we proceed similarly as in Ref (8), starting with the expression\cite
{yosh}

\begin{eqnarray}
E^{(2)} &=&\sum_{i}\langle \Psi ^{HF}\mid (H-H_{HF})\mid \Psi _{i}\rangle
\frac{1}{E^{HF}-E_{i}}\langle \Psi _{i}\mid (H-H_{HF})\mid \Psi ^{HF}\rangle
\label{foryosh} \\
&=&\sum_{i}\frac{\mid \langle \Psi ^{HF}\mid H\mid \Psi _{i}\rangle \mid ^{2}%
}{E^{HF}-E_{i}}.  \nonumber
\end{eqnarray}
Here $\Psi ^{HF}$, $E^{HF}$ are the mean field Slater determinant and total
Hartree-Fock energy associated with the ground state, respectively, and $H$
is the projection of the exact many particle Hamiltonian onto the first
Landau level. The many particle excited states $\Psi _{i}$ are Slater
determinants constructed with the basis (\ref{solution}), mixing states in
the filled band with states in the empty bands. It follows that $\langle
\Psi ^{HF}\mid \Psi _{i}\rangle =0$, a property that allowed to write the
last equality in (\ref{foryosh}). In the second quantized representation the
Hamiltonian $H$ will have non-vanishing matrix elements linking the HF state
and excited states of the form $\mid \Phi _{i}\rangle =a_{\mathbf{\eta }%
}\,a_{\mathbf{\eta }^{\prime }}\,a_{\mathbf{\xi }}^{+}\,a_{\mathbf{\xi }%
^{\prime }}^{+}\mid \Phi ^{HF}\rangle $ only, where $a_{\mathbf{\xi }}^{+}$
creates an electron of quantum numbers $\mathbf{\xi }$, etc. The index $i$
is a shorthand notation for the set of two pairs of filled $(\mathbf{\eta },%
\mathbf{\eta }^{\prime }\in F)$ and empty ($\mathbf{\xi },\mathbf{\xi }%
^{\prime }\in T-F$) electron states, where $T$ and $F\ $are the set of all
states in the Landau level and the filled ones, respectively. The indices $%
\mathbf{\eta }$ $\mathbf{=}(0,\mathbf{p,\sigma )}$ and $\mathbf{\xi =}(b%
\mathbf{,p,\sigma )}$ for $b=1,2$ denote the quantum numbers of filled and
empty states, respectively$.$ The total energy of an excited state is $%
E_{i}=E^{(HF)}+\epsilon (\mathbf{\xi })+\epsilon (\mathbf{\xi }^{\prime
})-\epsilon (\mathbf{\eta })-\epsilon (\mathbf{\eta }^{\prime })$. Then, the
second order correction can be rewritten in the form \cite{cabclaper}

\begin{equation}
E^{(2)}=\sum_{(\mathbf{\eta },\mathbf{\eta }^{\prime })}\sum_{(\mathbf{\xi },%
\mathbf{\xi }^{\prime })}\frac{\mid \langle \Phi ^{HF}\mid H\,a_{\mathbf{%
\eta }}\,a_{\mathbf{\eta }^{\prime }}\,a_{\mathbf{\xi }}^{+}\,a_{\mathbf{\xi
}^{\prime }}^{+}\mid \Phi ^{HF}\rangle \mid ^{2}}{{\Large \epsilon }(\mathbf{%
\eta })+{\Large \epsilon }(\mathbf{\eta }^{\prime })-{\Large \epsilon }(%
\mathbf{\xi })-{\Large \epsilon }(\mathbf{\xi }^{\prime })},
\end{equation}
where the total projected Hamiltonian is
\begin{eqnarray}
H &=&\frac{e^{2}}{2\varepsilon _{o}\,}\int \int d\mathbf{x}d\mathbf{x}%
^{\prime }\,\,\Psi ^{*}(\mathbf{x})\Psi ^{*}(\mathbf{x}^{\prime })\,\frac{1}{%
\mid \mathbf{x}-\mathbf{x}^{\prime }\mid }\,\Psi (\mathbf{x}^{\prime })\Psi (%
\mathbf{x}) \\
&=&\frac{e^{2}}{2\varepsilon _{o}}\sum_{\mathbf{\alpha },\mathbf{\alpha }%
^{\prime }}\sum_{\mathbf{\beta },\mathbf{\beta }^{\prime }}\,M(\mathbf{%
\alpha },\mathbf{\alpha }^{\prime }\mid \mathbf{\beta }^{\prime },\mathbf{%
\beta })\,\,\,\,a_{\mathbf{\alpha }}^{+}\,a_{\mathbf{\alpha }^{\prime
}}^{+}\,a_{\mathbf{\beta }^{\prime }}\,a_{\mathbf{\beta }}\,.  \nonumber
\end{eqnarray}
The matrix elements of the Coulomb interaction are given by
\begin{equation}
M(\mathbf{\alpha },\mathbf{\alpha }^{\prime }\mid \mathbf{\beta }^{\prime },%
\mathbf{\beta })\,=\int \int d\mathbf{x}d\mathbf{x}^{\prime }\,\,\Phi _{%
\mathbf{\alpha }}^{*}(\mathbf{x})\Phi _{\mathbf{\alpha }^{\prime }}^{*}(%
\mathbf{x}^{\prime })\,\frac{1}{\mid \mathbf{x}-\mathbf{x}^{\prime }\mid }%
\,\Phi _{\mathbf{\beta }^{\prime }}(\mathbf{x}^{\prime })\Phi _{\mathbf{%
\beta }}(\mathbf{x}),  \label{matrix}
\end{equation}
where we have used the shorthand notation $\Phi _{\mathbf{\alpha }}=$ $\Phi
_{\mathbf{p}}^{(b,\sigma )}$. By using the anti-commutation relations $[a_{%
\mathbf{\alpha }},a_{{\alpha }\mathbf{^{\prime }}}^{+}]=\delta _{\alpha
\mathbf{,\alpha }^{\prime }}$, formula (\ref{foryosh}) can be expressed as

\begin{equation}
E^{(2)}=\frac{e^{4}}{\varepsilon _{o}^{2}}\sum_{(\mathbf{\eta ,\eta }%
^{\prime })}\sum_{(\mathbf{\xi },\mathbf{\xi }^{\prime })}\frac{\mid \int
\int d\mathbf{x}d\mathbf{x}^{\prime }\,\Phi _{\mathbf{\eta },\mathbf{\eta }%
^{\prime }}^{*}(\mathbf{x},\mathbf{x}^{\prime })\frac{1}{\mid \mathbf{x}-%
\mathbf{x}^{\prime }\mid }\Phi _{\mathbf{\xi },\mathbf{\xi }^{\prime }}(%
\mathbf{x},\mathbf{x}^{^{\prime }})\mid ^{2}}{{\Large \epsilon }(\mathbf{%
\eta })+{\Large \epsilon }(\mathbf{\eta }^{\prime })-{\Large \epsilon }(%
\mathbf{\xi })-{\Large \epsilon }(\mathbf{\xi }^{\prime })},  \label{e2}
\end{equation}
where the two particle states $\Phi _{\mathbf{\eta },\mathbf{\eta }^{\prime
}}$ are defined by
\begin{equation}
\Phi _{\mathbf{\eta },\mathbf{\eta }^{\prime }}(\mathbf{x},\mathbf{x}%
^{\prime })=\frac{\Phi _{\mathbf{\eta }}(\mathbf{x})\Phi _{_{\mathbf{\eta }%
^{\prime }}}(\mathbf{x}^{\prime })-\Phi _{_{\mathbf{\eta }^{\prime }}}(%
\mathbf{x})\Phi _{\mathbf{\eta }}(\mathbf{x}^{\prime })}{\sqrt{2}}.
\end{equation}
The pairs $(\mathbf{\eta ,\eta }^{\prime })$and $(\mathbf{\xi },\mathbf{\xi }%
^{\prime })$ are considered as unordered.

\subsection{Correlation energy: first evaluation}

As pointed out above the single particle bands are remarkably flat. We can
use this property to simplify the calculation of the energy correction.
First, we approximate the filled band energies appearing in the denominator
of (\ref{e2}) by their mean value,
\[
\epsilon (\eta )=\epsilon _{0}.
\]
In addition, and slightly more crudely, we substitute the energies in the
excited bands by a common energy equal to half the sum of the mean energies
of the two bands,
\[
\epsilon (\xi )=\frac{\epsilon _{1}+\epsilon _{2}}{2}.
\]
This last approximation is taken in view of the small relative gap
separating these two bands.

With this simplifying substitution (\ref{e2}) can be expressed in the
simpler form
\begin{eqnarray}
E^{(2)} &=&\frac{e^{4}}{2(2\epsilon _{0-}\epsilon _{1-}\epsilon
_{2})\varepsilon _{o}^{2}}\int \int \,d\mathbf{x}^{\prime }d\mathbf{x}\int
\int d\mathbf{y}d\mathbf{y}^{\prime }\frac{1}{\mid \mathbf{x}-\mathbf{x}%
^{\prime }\mid }\frac{1}{\mid \mathbf{y}-\mathbf{y}^{\prime }\mid }\times  \\
&&\left( \pi _{f}(\mathbf{x}^{\prime },\mathbf{y}^{\prime })\pi _{f}(\mathbf{%
x},\mathbf{y})-\pi _{f}(\mathbf{x}^{\prime },\mathbf{y})\pi _{f}(\mathbf{x},%
\mathbf{y}^{\prime })\right) \pi _{e}(\mathbf{y},\mathbf{x})\pi _{e}(\mathbf{%
y}^{\prime },\mathbf{x}^{\prime }),  \nonumber
\end{eqnarray}
where $\pi _{f}$ is the projection operator on the subspace of states of the
filled band and $\pi _{e}$ the projector associated to the subspace of
states formed with the union of the empty bands. The projectors have the
following expression in terms of the corresponding densities
\begin{eqnarray}
\pi _{f}(\mathbf{x},\mathbf{x}^{\prime }) &=&\sum_{\mathbf{p},\sigma }\Phi _{%
\mathbf{p}}^{(0,\sigma )}(\mathbf{x})(\Phi _{\mathbf{p}}^{(0,\sigma )}(%
\mathbf{x}^{\prime }))^{*}, \\
&=&P(\mathbf{x},\mathbf{x}^{\prime })2\pi r_{o}^{2}\sum_{Q}\rho _{f}(\mathbf{%
Q})\exp (i\mathbf{Q}.(\frac{\mathbf{x+x}^{\prime }}{2}+\frac{1}{2}i\ \mathbf{%
n\times (x-x}^{\prime }\mathbf{)})),  \nonumber \\
\pi _{e}(\mathbf{x},\mathbf{x}^{\prime }) &=&\sum_{\mathbf{p},b,\sigma }\Phi
_{\mathbf{p}}^{(b,\sigma )}(\mathbf{x})(\Phi _{\mathbf{p}}^{(b,\sigma )}(%
\mathbf{x}^{\prime }))^{*},  \nonumber \\
&=&P(\mathbf{x},\mathbf{x}^{\prime })2\pi r_{o}^{2}\sum_{Q}\rho _{t}(\mathbf{%
Q})\exp (i\mathbf{Q}.(\frac{\mathbf{x+x}^{\prime }}{2}+\frac{1}{2}i\ \mathbf{%
n\times (x-x}^{\prime }\mathbf{)})),  \nonumber \\
\rho _{e}(\mathbf{Q}) &=&\frac{1}{2\pi r_{o}^{2}}\delta _{\mathbf{Q,0}}-\rho
_{b}(\mathbf{Q}),  \nonumber
\end{eqnarray}
where $\delta _{\mathbf{Q},\mathbf{0}}$ is the ordinary Kronecker delta and $%
P(\mathbf{x},\mathbf{x}^{\prime })_{\text{ }}$is the projection operator
onto the first Landau level, defined in Appendix A. These expressions can be
obtained from formula (\ref{onepart}) in Appendix B. After evaluating a few
spatial integrals, the following formula for the correlation energy per
particle $\epsilon ^{(2)}=E^{(2)}/N$ is obtained
\begin{eqnarray}
\epsilon ^{(2)} &=&\frac{e^{4}}{2(2\epsilon _{0-}\epsilon _{1-}\epsilon
_{2})\varepsilon _{o}^{2}}\frac{4\pi \sqrt{\pi }r_{o}(\pi r_{o}^{2})^{2}}{%
\nu }\sum_{\mathbf{Q}_{1}}\sum_{\mathbf{Q}_{2}}\sum_{Q_{3}}\rho _{f}(\mathbf{%
Q}_{1})\rho _{f}(\mathbf{Q}_{2})\rho _{e}(\mathbf{Q}_{3})\rho _{e}(\mathbf{-Q%
}_{1}\mathbf{-Q}_{2}\mathbf{-Q}_{3}) \\
&&\exp (\frac{(\mathbf{Q}_{1}\mathbf{+Q}_{2})^{2}r_{o}^{2}}{4})\int \mathbf{%
dz}\frac{1}{\sqrt{\mathbf{z}^{2}}}\exp (-\frac{\mathbf{z}^{2}}{4r_{o}^{2}}%
)\exp \left( -(\mathbf{Q}_{1}\mathbf{+Q}_{3}).\frac{\mathbf{n\times z}}{2}-i(%
\mathbf{Q}_{2}\mathbf{+Q}_{3}).\frac{\mathbf{z}}{2}\right)   \nonumber \\
&&I_{0}\left( \frac{1}{8}(\frac{\mathbf{z}}{r_{o}}-\mathbf{n\times (Q}_{1}%
\mathbf{+Q}_{3}\mathbf{)}r_{o}-i(\mathbf{Q}_{2}\mathbf{+Q}_{3}\mathbf{)}%
r_{o})^{2}\right) \times   \nonumber \\
&&(\exp \frac{1}{8}(\frac{\mathbf{z}}{r_{o}}-\mathbf{n\times (Q}_{1}\mathbf{%
+Q}_{3}\mathbf{)}r_{o}-i(\mathbf{Q}_{2}\mathbf{+Q}_{3})r_{o})^{2}-\exp (-%
\frac{1}{8}(\frac{\mathbf{z}}{r_{o}}-\mathbf{n\times (Q}_{1}\mathbf{+Q}_{3}%
\mathbf{)}r_{o}-i\mathbf{(Q}_{2}\mathbf{+Q}_{3}\mathbf{)}r_{o})^{2})).
\nonumber
\end{eqnarray}

Further progress in the evaluation of the energy correction is aided by
noting the presence of an exponential factor in the squared Fourier
wave-vectors, allowing to take just a few of them only, for sufficient
convergence. We thus assume the 36 Fourier components associated with the
shortest values of $\mathbf{Q}$ are nonzero, only. Of these, those
associated with the longest $\mathbf{Q}$ are slightly altered to assure that
the sum rule\cite{clar2}
\begin{equation}
\sum_{Q}|\rho _{f}(\mathbf{Q})|^{2}\exp (\frac{\mathbf{Q}^{2}r_{o}^{2}}{2})=%
\frac{\nu }{(2\pi r_{o}^{2})^{2}}
\end{equation}
is satisfied. This condition is used to cancel various fictitious
divergences in the formula defining the correlation energy \cite{yosh}.
Then, we obtain

\begin{eqnarray}
\epsilon ^{(2)} &=&\frac{(e^{2}/\varepsilon _{o}r_{0})^{2}}{4\ \nu \sqrt{\pi
}\ (2\epsilon _{0-}\epsilon _{1-}\epsilon _{2})}\sum_{\mathbf{Q}_{1}}\sum_{%
\mathbf{Q}_{2}}\sum_{\mathbf{Q}_{3}}\Delta _{f}(\mathbf{Q}_{1})\Delta _{f}(%
\mathbf{Q}_{2})\Delta _{e}(\mathbf{Q}_{3})\Delta _{e}(\mathbf{Q}_{1}+\mathbf{%
Q}_{2}+\mathbf{Q}_{3})\times   \label{correc1} \\
&&\exp (-i\ \frac{r_{o}}{2}\mathbf{(Q_{2}+Q_{3})\cdot n\times (Q_{1}+Q_{3})}%
r_{o}+\frac{(\mathbf{Q}_{2}\mathbf{+Q}_{3})^{2}r_{o}^{2}}{2})\times
\nonumber \\
&&\int \mathbf{du}\ \exp \left( -\frac{(\mathbf{u-i(Q_{2}+Q_{3})}r_{o}%
\mathbf{)}^{2}}{4}-i\ r_{o}(\mathbf{Q}_{2}\mathbf{+Q}_{3}).\frac{\mathbf{%
(u-i(Q_{2}+Q_{3})}r_{o}\mathbf{)}}{2}\right) \times   \nonumber \\
&&I_{0}\left( \frac{1}{8}(\mathbf{u}-i(\mathbf{Q}_{2}\mathbf{+Q}_{3}\mathbf{)%
}r_{o})^{2}\right) \sinh \left( \frac{1}{8}(\mathbf{u}-i(\mathbf{Q}_{2}%
\mathbf{+Q}_{3})r_{o})^{2}\right) \times   \nonumber \\
&&\left( \frac{1}{\sqrt{(\mathbf{u+n\times (Q_{1}+Q_{3})}r_{o}\mathbf{)}^{2}}%
}-\frac{{\Large \delta }_{\mathbf{Q}_{2}+\mathbf{Q}_{3},\mathbf{0}}}{\sqrt{%
\mathbf{u}^{2}}}\right) ,  \nonumber
\end{eqnarray}
where the order parameters of the filled and empty bands have been defined
as usual,
\begin{eqnarray*}
\Delta _{f}(\mathbf{Q}) &=&2\pi r_{o}^{2}\rho _{f}(\mathbf{Q})\exp (\frac{(%
\mathbf{Q})^{2}r_{o}^{2}}{4}), \\
\Delta _{e}(\mathbf{Q}) &=&2\pi r_{o}^{2}\rho _{e}(\mathbf{Q})\exp (\frac{(%
\mathbf{Q})^{2}r_{o}^{2}}{4}).
\end{eqnarray*}

\begin{figure}[tbp]
\includegraphics[width=3.5in]{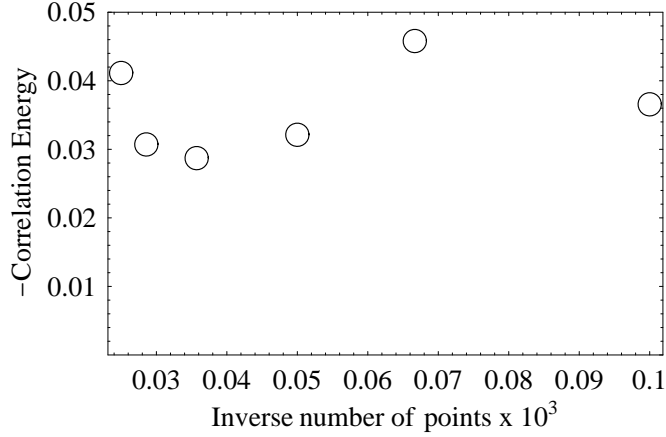}
\caption{Second order correction to the energy per particle as a function of
inverse number of sample points in the Monte-Carlo integration.}
\label{fig:data1}
\end{figure}

The integrals in Eq. (\ref{correc1}) were calculated using the Monte-Carlo
method with a variable number of sample points in the range $N=10000-40000$.
The region of integration was taken as a square of size $L=50$ in units of
the magnetic length $r_{0},$ centered at the origin. At 1/3 filling this
sample contains 132 particles and traps 396 flux quanta. The results for the
correlation energy as a function of $1/N$ are shown in Fig.~\ref{fig:data1}.
In units $e^{2}/\varepsilon _{o}r_{o}$ the estimated value for the
correlation energy is
\begin{equation}
\epsilon ^{(2)}\simeq -0.04\pm 0.01\ .  \label{e2num1}
\end{equation}
Adding this correction to the Hartree-Fock energy (\ref{energy}) one obtains
the value $\varepsilon \simeq -0.40\pm 0.01$, to be compared with the result
of Yoshioka and Lee with its own correction $\varepsilon \simeq -0.394$, and
that of Laughlin $\varepsilon \simeq -0.416$, all in the same units.

\subsection{ Correlation energy: dependence on the sample size}

The lack of a definite convergence pattern provided by the method just
described prompted us to try a different procedure. Using the formulas
derived in Appendix D it is possible to recast Eq. (\ref{correc1}) to give
it the form,

\begin{eqnarray}
\epsilon ^{(2)} &=&\frac{3(\frac{e^{2}}{\varepsilon _{o}r_{0}})^{2}}{%
(N_{\phi _{0}})^{3}}\sum_{(p_{\beta },\ \sigma _{\beta })}\sum_{(p_{\beta
^{\prime }},\ \sigma _{\beta ^{\prime }})}\sum_{(b,\ p_{\alpha },\ \sigma
_{\alpha })}\sum_{(b^{\prime },p_{\alpha ^{\prime }},\sigma _{\alpha
^{\prime }})}  \label{e2alter} \\
&&\frac{1}{\epsilon ^{(0,\ p_{\beta },\ \sigma _{\beta })}+\epsilon ^{(0,\
p_{\beta ^{\prime }},\ \sigma _{\beta ^{\prime }})}-\epsilon ^{(b,\
p_{\alpha },\ \sigma _{\alpha })}-\epsilon ^{(b^{\prime },\ p_{\alpha
^{\prime }},\ \sigma _{\alpha ^{\prime }})}}\times   \nonumber \\
&\mid &\sum_{r_{\alpha },r_{\alpha ^{\prime }},r_{\beta },r_{\beta ^{\prime
}}}\delta ^{(K)}(\mathbf{P}_{\beta ^{\prime },r_{\beta ^{\prime }}}+\mathbf{P%
}_{\beta ,r_{\beta }}-\mathbf{P}_{a,r_{\alpha }}\mathbf{-P}_{\alpha ^{\prime
},r_{\alpha ^{\prime }}},0)\times   \nonumber \\
&&\mathcal{F}^{*}(\mathbf{p}_{\alpha },r_{\alpha },\sigma _{\alpha })\
g_{r_{\alpha }}^{*b}(\mathbf{p}_{\alpha })\mathcal{F}^{*}(\mathbf{p}_{\alpha
^{\prime }},r_{\alpha ^{\prime }},\sigma _{\alpha ^{\prime }})\ g_{r_{\alpha
^{\prime }}}^{*b^{\prime }}(\mathbf{p}_{\alpha ^{\prime }})\mathcal{F}(%
\mathbf{p}_{\beta ^{\prime }},r_{\beta ^{\prime }},\sigma _{\beta ^{\prime
}})\ g_{r_{\beta ^{\prime }}}^{0}(\mathbf{p}_{\beta ^{\prime }})\times
\nonumber \\
&&\mathcal{F}(\mathbf{p}_{\beta },r_{\beta },\sigma _{\beta })\ g_{r_{\beta
}}^{0}\text{(}\mathbf{p}_{\beta }\text{\ )}\mathcal{F}_{\mathbf{P}%
_{a,r_{\alpha }}}^{*}(\mathbf{n\times (P}_{\beta ,r_{\beta }}\mathbf{-P}%
_{a,r_{\alpha }}\mathbf{)}r_{0}^{2})\times   \nonumber \\
&&\mathcal{F}_{\mathbf{P}_{\alpha ^{\prime },r_{\alpha ^{\prime }}}}^{*}(%
\mathbf{n\times (P}_{\beta ^{\prime },r_{\beta ^{\prime }}}\mathbf{-P}%
_{\alpha ^{\prime },r_{\alpha ^{\prime }}}\mathbf{)}r_{0}^{2})\ \times
V(\alpha ,\alpha ^{\prime },\beta ,\beta ^{\prime })\ |^{2},\hspace{1.04in}
\nonumber
\end{eqnarray}
where the function $V$ is defined by

\begin{eqnarray*}
V(\alpha ,\alpha ^{\prime },\beta ,\beta ^{\prime }) &=&\sum_{\mathbf{Q}^{*}}%
\frac{2\pi }{r_{0}\mid \mathbf{P}_{\beta ^{\prime },r_{\beta ^{\prime }}}%
\mathbf{-P}_{\alpha ^{\prime },r_{\alpha ^{\prime }}}+\mathbf{Q}^{*}\mid }%
\exp (-r_{0}^{2}(\mathbf{P}_{\beta ^{\prime },r_{\beta ^{\prime }}}\mathbf{-P%
}_{\alpha ^{\prime },r_{\alpha ^{\prime }}}+\mathbf{Q}^{*})^{2})\times \  \\
&&\exp \left[ -i\ r_{0}^{2}\ \mathbf{n}\times \mathbf{Q}^{*}\cdot (\mathbf{P}%
_{\alpha ,r_{\alpha }}+\mathbf{P}_{\beta ,r_{\beta }}-\mathbf{P}_{\beta
^{\prime },r_{\beta ^{\prime }}}\mathbf{-P}_{\alpha ^{\prime },r_{\alpha
^{\prime }}})\right] ,
\end{eqnarray*}
The form of the pure phase factors $\mathcal{F}^{*}$ and $\mathcal{F}_{%
\mathbf{P}}^{*}$ , the momenta $\mathbf{P}_{\alpha ,r}$ and the special
reciprocal lattice vectors $\mathbf{Q}^{*}$ are all specified in Appendix D.
As before, the functions $g_{r}^{b}(\mathbf{p})$ are the coefficients
determining the single particle HF excitations. We remark that once the
coefficients for the filled band are known the other set of coefficients for
the empty bands can be evaluated analytically as the two eigenvectors of the
(3x3) matrix representation of the single particle HF hamiltonian,
orthogonal to the vector $g_{r}^{0}(\mathbf{p})$.

\begin{figure}[tbp]
\includegraphics[width=3.5in]{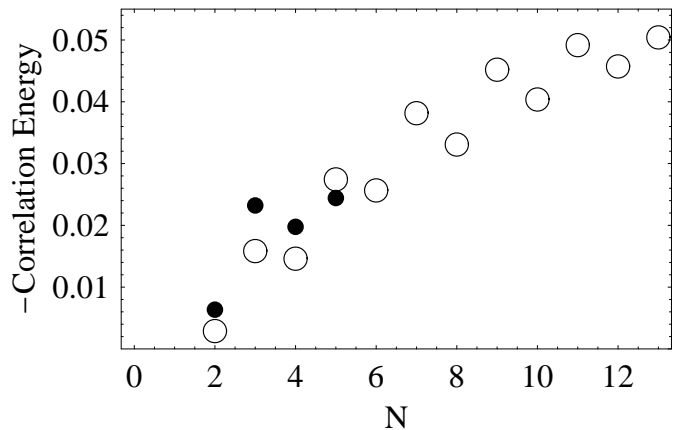}
\caption{Correlation energy as a function of the parameter $N$. The circles
indicate the calculated values for elongated samples of the form $%
(N_{1},N_{2})=(1,N)$ . The black dots signal the values corresponding more
appropriate samples with form defined by $(N_{1},N_{2})=(N,N).$}
\label{fig:data2}
\end{figure}

\bigskip \ In order to evaluate numerically expression (\ref{e2alter}) we
restrict the Hilbert space of the single particle HF\ problem by
defining a cell of sides $2N_{1}\mathbf{a}_{1}$,
$2N_{2}\mathbf{a}_{2}$ with $N_{1}$, $N_{2}$ integers, and
imposing periodic boundary conditions over its borders. Using the
symmetry properties (\ref{symmetry})

\begin{eqnarray*}
T_{N_{1}6\mathbf{c}_{1}}\chi _{\mathbf{p}}^{(r,\sigma )}(\mathbf{x}) &=&\exp
(-i6\ \mathbf{p.c}_{1}N_{1})\chi _{\mathbf{p}}^{(r,\sigma )}(\mathbf{x}), \\
T_{N_{2}6\mathbf{c}_{2}}\chi _{\mathbf{p}}^{(r,\sigma )}(\mathbf{x}) &=&\exp
(-i6\ \mathbf{p.c}_{2}N_{2})\chi _{\mathbf{p}}^{(r,\sigma )}(\mathbf{x}).
\end{eqnarray*}
these constraints can be translated into the relations

\begin{eqnarray*}
-6\mathbf{p.c}_{1} &=&2\pi \frac{n_{1}}{N_{1}}, \\
-6\mathbf{p.c}_{2} &=&2\pi \frac{n_{2}}{N_{2}}.
\end{eqnarray*}
They restrict the values of the quasimomentum $\mathbf{p}$ to a discrete
set, as expressed in compact form by the condition
\begin{eqnarray}
\mathbf{p} &=&\mathbf{-}\frac{n_{1}}{N_{1}}\frac{\mathbf{t}_{1}}{2}-\frac{%
n_{2}}{N_{2}}\frac{\mathbf{t}_{2}}{3},  \label{restrict} \\
0 &\leq &n_{1}<N_{1}\text{ \ },0\leq n_{2}<N_{2},  \nonumber
\end{eqnarray}
where $\mathbf{t}_{1}$ and $\mathbf{t}_{2}$ are defined in Appendix A. The
cell chosen traps $6N_{1}N_{2}$ flux quanta and contains $2N_{1}N_{2}$
particles. The integrals over the continuum in the Brillouin zone of Eq. (%
\ref{e2alter}) are now restricted to summations over a finite number of
values of $\mathbf{p}$. Formula (\ref{e2alter}) was first evaluated by for
an elongated sample with \ $N_{1}=1,$ and $N_{2}=N\ $ varying from $1$ to $13
$. \ In order to keep the computing time within our allowed practical
limits, the values of $\ n_{1}$ and $n_{2}$ in the expression defining the
summation argument
\begin{equation}
\mathbf{Q}^{*}=n_{1}\frac{3}{2}\ \mathbf{t}_{1}+n_{2}\frac{2}{3}\ \mathbf{t}%
_{2},
\end{equation}
were restricted to the regions $-4\leq n_{1}\leq 4$ and $-4\leq n_{2}\leq 4.$
The results of this calculation are shown as empty circles in Fig.~(\ref
{fig:data2}). Note that the correlation energy per particle grows with N,
approaching rather slowly the thermodynamic limit whose value $\epsilon
^{(2)}\sim -0.06\ e^{2}/\epsilon _{o}r_{o}$ exceeds the result (\ref{e2num1}%
) of the previous method. The energy correction was also calculated for
samples with an aspect ratio close to one, taking $(N_{1},N_{2})=(N,N)$, up
to $N=5,$ an upper bound set by computing limitations. The results are shown
in Fig.~(\ref{fig:data2}) by filled circles. \ As can be observed,\ the
energy values for those samples grow similarly with $N$ than those
associated with elongated samples. This property indicates that these more
appropriate samples also need values of $N$ greater than five in order to
approach convergence. However, already at $N=5$ the number of particles
involved is $2N^{2}=50$, considerably higher than the numbers handled in
exact diagonalization.

The most relevant finding of this last method of evaluation is the need for
very high sample sizes in order to approach the thermodynamical limit. This
outcome strongly supports the relevance of \ long distance correlations in
the attainment of the true character of the ground state through numerical
computations.

\section{Summary}

An analytic solution of the Hartree-Fock problem for a 2DEG at filling $1/3$
and 1/2 electron per unit cell is found by employing symmetry considerations
and a special complete set of common eigenfunctions of the magnetic
translations. A triplet of bands arises that turn out surprisingly flat as
functions of the two dimensional quasimomentum. The Coulomb interaction
breaks the first Landau level in three narrow sub-bands resembling effective
Landau levels of composite fermions, one of which is fully occupied and the
other two, empty. The energy per particle reproduces an earlier numerical
result for this quantity \cite{clar2}. We find that the charge density forms
hexagonal rings throughout the lattice, suggesting strong long range
correlations owing to cooperative ring exchange. An evaluation of the
correlation energy using second order perturbation theory yields a
correction an order of magnitude larger than that for the localized
single-particle features of the usual Wigner Crystal state,\cite{yosh} and
lowers the energy to make it comparable with the lowest values obtained
using other models. Our results also suggest that the thermodynamic limit in
a cell with aspect ratio near one is not reached if less than 50 particles
are included.

\section{Acknowledgments}

This work was supported in part by Fondecyt Grants 1020829 and 7020829, the
ICTP Associateship Program, and the Third World Academy of Sciences. Helpful
discussions with A. Gonzalez, G. Baskaran, N. H March and K. Esfarjani are
also acknowledged.

\section{Appendix A}

\subsection{Eigenfunctions of magnetic translations}

Consider a 2DEG constrained to move in a plane of area $A$ under a
perpendicular magnetic field $\mathbf{B}$. A useful basis set of single
particle Bloch-like states in the lowest Landau level can be defined in
terms of linear combinations of the normalized zero angular momentum
eigenfunction

\begin{equation}
\phi (\mathbf{x})=\frac{1}{\sqrt{2\pi }r_{o}}\exp (-\frac{\mathbf{x}^{2}}{%
4r_{o}^{2}})
\end{equation}
in the compact form \cite{wannier,ferrari,cabo0,cabo1},

\begin{equation}
\varphi _{k}(\mathbf{x})=\frac{1}{N_{\mathbf{k}}}\sum_{\mathbf{\ell }}\left(
-1\right) ^{\ell _{1}\ell _{2}}\exp (i\ \mathbf{k}.\mathbf{\ell )\ }T\mathbf{%
_{{\ell }}\ \phi (x),}  \label{fi}
\end{equation}
\[
N_{\mathbf{k}}=\sqrt{N_{\phi _{0}}}\sqrt{\sum_{\mathbf{\ell }}\left(
-1\right) ^{\ell _{1}\ell _{2}}\exp (i\ \mathbf{k}.\mathbf{\ell -}\frac{%
\mathbf{\ell }^{2}}{4r_{0}^{2}}\mathbf{)\ .}}
\]
Due to its role in the above definition the function $\phi $ is called the
''seed'' function. The sum runs over all integers $\ell _{1},\ell _{2}$
defining a planar lattice $L$, through $\mathbf{\ell }=\ell _{1}\mathbf{b}%
_{1}+\ell _{2}\mathbf{b}_{2},$ where the unit cell intercepts one flux
quantum, so that $\mathbf{n.b}_{1}\mathbf{\times b}_{2}=2\pi r_{o}^{2}.$ \
The magnetic translation operators $T\mathbf{_{R}}$ acting on any function $f
$ are defined by
\begin{equation}
T_{\mathbf{R}}\mathbf{\ }f(\mathbf{x})\mathbf{=}\exp \mathbf{(}\frac{2ie}{%
\hbar c}\mathbf{A({R}).x)\ }f(\mathbf{x-R})\mathbf{,}  \label{magntra}
\end{equation}
where the vector potential is assumed in the axial gauge $\mathbf{A}(\mathbf{%
x})=B(-x_{2},x_{1},0)/2$ and the electron charge $e$ is taken with its
negative sign. This basis was employed before to obtain exact mean field
solutions of the related problem at filling 1/2.\cite{cabclaper} For
arbitrary vectors $\mathbf{R}_{1}$ and $\mathbf{R}_{2}$ the translation
operators satisfy the commutation relation
\begin{equation}
T_{\mathbf{R}_{1}}T_{\mathbf{R}_{2}}=\exp (\frac{ie}{\hbar c}\mathbf{A(R}_{1}%
\mathbf{).R}_{2})T_{\mathbf{R}_{2}}T_{\mathbf{R}_{1}}.  \label{commutation}
\end{equation}
As it may be easily verified, the functions $\varphi _{\mathbf{k}}\ $
satisfy the eigenvalue equation

\begin{eqnarray}
T_{\mathbf{\ell }}\ \varphi _{\mathbf{k}}(\mathbf{x}) &=&\lambda _{\mathbf{k}%
}(\mathbf{\ell })\ \varphi _{\mathbf{k}}(\mathbf{x}),  \label{eigen1} \\
\lambda _{\mathbf{k}}(\mathbf{\ell }) &=&\left( -1\right) ^{\ell _{1}\ell
_{2}}\exp (-i\ \mathbf{k}.\mathbf{\ell }).\   \label{eigen2}
\end{eqnarray}
Arranged in an arbitrary Slater determinant these functions are exact
solutions of the Hartree-Fock problem.\cite{cabo0,cabo2} This strong
property happens because the HF single particle Hamiltonian commutes with
all translations leaving $L$ invariant. \cite{cabo0} The functions (\ref{fi}%
) are common eigenfunctions of the commuting magnetic translations.\
Moreover, the set of eigenvalues (\ref{eigen2}) uniquely determines them.
Therefore, the HF hamiltonian associated with the Slater determinant can not
change those eigenvalues and the $\varphi _{\mathbf{k}}$ should be
eigenfunctions.

Finally, let us show that the effect of an arbitrary translation on the
basis functions is equivalent to a shift in the momentum label, modulo a
phase factor.\cite{ferrari} Operating twice with the translation operator
involving an arbitrary vector $\mathbf{a}$ and a vector in the lattice $%
\mathbf{\ell }$ and using Eqs. (\ref{commutation}) and (\ref{eigen1}) one
readily gets,

\begin{eqnarray}
T_{\mathbf{a}}T_{\mathbf{\ell }}\ \varphi _{\mathbf{p}}(\mathbf{x})
&=&\lambda _{\mathbf{p}}(\mathbf{\ell )}T_{\mathbf{a}}\varphi _{\mathbf{p}}(%
\mathbf{x})  \label{transl1} \\
&=&\exp (2\frac{ie}{\hbar c}\mathbf{A}(\mathbf{a}).\mathbf{\ell })\ T_{\ell
}T_{\mathbf{a}}\ \varphi _{\mathbf{p}}(\mathbf{x}).  \nonumber
\end{eqnarray}
Again using (\ref{eigen1}) we have

\[
T_{\mathbf{\ell }}T_{\mathbf{a}}\ \varphi _{\mathbf{p}}(\mathbf{x})=\lambda
_{\mathbf{p}+2\frac{e}{\hbar c}\mathbf{A}(\mathbf{a})}(\mathbf{\ell })T_{%
\mathbf{a}}\ \varphi _{\mathbf{p}}(\mathbf{x}).
\]
Then, taking into account that the set of eigenvalues defines uniquely the
wave-functions modulo a phase, it follows that

\begin{equation}
T_{\mathbf{a}}\ \varphi _{\mathbf{p}}(\mathbf{x})=\mathcal{F}_{\mathbf{p}}(%
\mathbf{a})\ \varphi _{\mathbf{p}+2\frac{e}{\hbar c}\mathbf{A}(\mathbf{a})}(%
\mathbf{x}),  \label{phase}
\end{equation}
from which also follows,

\begin{equation}
\mathcal{F}_{\mathbf{p}}(\mathbf{a})=\frac{\varphi _{\mathbf{p}}(\mathbf{0})%
}{\varphi _{\mathbf{p}+2\frac{e}{\hbar c}\mathbf{A}(\mathbf{a})}(\mathbf{a})}%
.  \label{fase1}
\end{equation}
That is, a magnetic translation is equivalent to a shift in the
quasi-momentum.

\subsection{A $\varphi _{\mathbf{p}}$-transform}

Any function$\ f$ in the first Landau level, and its inverse, can be
represented as

\begin{eqnarray*}
f(\mathbf{x}) &=&\sum_{\mathbf{p}\in \tilde{B}}c(\mathbf{p})\ \varphi _{%
\mathbf{p}}(\mathbf{x}), \\
&=&\int_{\mathbf{p}\in \tilde{B}}\frac{d\mathbf{p}}{(2\pi )^{2}}\tilde{c}(%
\mathbf{p})\ \bar{\varphi}_{\mathbf{p}}(\mathbf{x}), \\
\tilde{c}(\mathbf{p}) &=&\int d\mathbf{x\ }\ \bar{\varphi}_{\mathbf{p}}^{*}(%
\mathbf{x})\ f(\mathbf{x}),
\end{eqnarray*}
where $\bar{\varphi}_{\mathbf{p}}(\mathbf{x})=\sqrt{S\ }\varphi _{\mathbf{p}%
}(\mathbf{x}),$ $\tilde{c}(\mathbf{p})=\sqrt{S}\ c(\mathbf{p}),$ with $%
S=2\pi r_{0}^{2}\ N_{\phi _{0}},$ with $N_{\phi _{0}}$ the number
of flux quanta in the system area, $\tilde{B}$ is the Brillouin
zone defined by the unit cell vectors
\begin{eqnarray*}
\mathbf{t}_{1} &=&-\frac{1}{r_{0}^{2}}\mathbf{n\times b}_{2}, \\
\mathbf{t}_{2} &=&\frac{1}{r_{0}^{2}}\mathbf{n\times b}_{1}.
\end{eqnarray*}

The orthogonality and completeness relations in the first Landau level of
the modified functions take the forms

\begin{eqnarray}
\int_{A}d\mathbf{x\ }\bar{\varphi}_{\mathbf{p}}^{*}(\mathbf{x})\bar{\varphi}%
_{\mathbf{p}^{\prime }}(\mathbf{x}) &=&(2\pi )^{2}\delta (\mathbf{p-p}%
^{\prime }),  \nonumber \\
P(\mathbf{x},\mathbf{x}^{\prime }) &=&\sum_{\mathbf{p}\in \tilde{B}}\varphi
_{\mathbf{p}}(\mathbf{x})\ \varphi _{\mathbf{p}}^{*}(\mathbf{x}^{\prime
})=\int_{p\in \tilde{B}}\frac{dp}{(2\pi )^{2}}\ \bar{\varphi}_{\mathbf{p}}(%
\mathbf{x})\ \bar{\varphi}_{\mathbf{p}}^{*}(\mathbf{x}^{\prime })
\end{eqnarray}

\subsection{The orbitals $\chi _{\mathbf{p}}^{(r,\sigma )}$ as special cases
of the basis functions $\varphi _{\mathbf{p}}(\mathbf{x})$\ }

Lets us now verify that the functions $\chi _{\mathbf{p}}^{(r,\sigma )}$
constructed to diagonalize in blocks the HF Hamiltonian in Sec. 2 are simply
constant phase factors multiplied by a particular kind of basis functions $%
\varphi _{\mathbf{p}}$. These differ from the orbitals considered in Sec. 2
and defined by the unit cell vectors (\ref{bas32}) in that their basis
vectors are given by
\begin{eqnarray}
\mathbf{b}_{1} &=&\frac{2}{3}\mathbf{a}_{1}, \\
\mathbf{b}_{2} &=&\mathbf{a}_{2}.  \nonumber
\end{eqnarray}
To start out it will be useful to consider the $\varphi _{\mathbf{p}}$%
-transform of \ a translated ''seed'' function$\ \phi (x).$ For this purpose
we notice that it is possible to fix the second argument of the projector
operator to be equal to the translation vector $\mathbf{a.}$ \ It follows
that the projector operator in the first Landau level can be rewritten as
\begin{eqnarray}
P(\mathbf{x},\mathbf{a}) &=&\frac{1}{2\pi r_{0}^{2}}\exp (-\frac{(\mathbf{x-a%
})^{2}}{4r_{0}^{2}})\exp (\frac{ie}{\hbar c}\mathbf{A(a).x}) \\
&=&\frac{1}{\sqrt{2\pi r_{0}^{2}}}T_{a}\phi (\mathbf{x})=\int_{p\in \tilde{B}%
}\frac{dp}{(2\pi )^{2}}\ \bar{\varphi}_{\mathbf{p}}(\mathbf{x})\ \bar{\varphi%
}_{\mathbf{p}}^{*}(\mathbf{a}),  \nonumber
\end{eqnarray}
where the last equality is merely the definition of the projector operator
in coordinate space. Therefore, these relations determine the following $%
\varphi _{\mathbf{p}}$-transform of the translated ''seed'' function
\begin{equation}
T_{a}\phi (\mathbf{x})=\int_{\mathbf{p}\in \tilde{B}}\frac{d\mathbf{p}}{%
(2\pi )^{2}}\ \left( \sqrt{2\pi r_{0}^{2}}\bar{\varphi}_{\mathbf{p}}^{*}(%
\mathbf{a})\right) \bar{\varphi}_{\mathbf{p}}(\mathbf{x}).\
\label{transform}
\end{equation}
Next we will consider the definition
\begin{equation}
\chi _{\mathbf{p}}^{(r,\sigma )}(x)=\frac{1}{\sqrt{6}N_{\mathbf{p}}^{(3,2)}}%
\sum_{\mathbf{m}}\exp (i\mathbf{P}^{(\mathbf{p},r,\sigma )}.\mathbf{m}+\frac{%
5\pi i}{6}m_{1}m_{2})T_{\mathbf{m}}\phi (\mathbf{x})\text{ ,\ }
\label{csinew}
\end{equation}
and represent the vectors $\mathbf{m}$\ in terms of alternative vectors $%
\mathbf{\ell }$ and the indices $\alpha $ and $\beta $ defined through the
expressions

\begin{eqnarray*}
\mathbf{m} &=&\mathbf{\ell }+\delta , \\
\mathbf{\ell } &=&\mathbf{\ell }_{1}2\ \mathbf{c}_{1}+\mathbf{\ell }_{2}\ 3\
\mathbf{c}_{2}, \\
\delta  &=&u\ \mathbf{c}_{1}+v\ \mathbf{c}_{2}.
\end{eqnarray*}
The symbols $\alpha ,\beta \mathbf{,\ \ell }_{1}$ and $\mathbf{\ell }_{2}$
are specified by the relations

\begin{eqnarray*}
u &=&\left\langle m_{1}\right\rangle ={\Huge \{}
\begin{tabular}{l}
$0\ \ \ if\ \ m_{1}=0\ \ Modulo(2)$ \\
$1\ \ \ if\ \ m_{1}=1\ \ Modulo(2)$%
\end{tabular}
, \\
v &=&\left[ m_{1}\right] ={\Huge \{}
\begin{tabular}{l}
$-1$ \ \ \ \ $\ if$ $\ \ m_{2}=-1$ $\ Modulo(3)$ \\
$\ \ 0$ \ \ \ \ \ $if$ $\ \ m_{2}=$ $\ \ \ 0$ $\ Modulo(3)$ \\
$\ \ 1$ \ \ \ \ \ $if$ $\ \ m_{2}=$ $\ \ \ 1$ $\ Modulo(3)$%
\end{tabular}
, \\
\mathbf{\ell }_{1} &=&\frac{m_{1}-\left\langle m_{1}\right\rangle }{2}, \\
\mathbf{\ell }_{2} &=&\frac{m_{2}-\left[ m_{2}\right] }{3\ }.
\end{eqnarray*}

The use of these alternative definitions allows to write (\ref{csinew}) in
the form

\begin{eqnarray}
\chi _{\mathbf{p}}^{(r,\sigma )}(x) &=&\frac{1}{\sqrt{6}N_{\mathbf{p}%
}^{(3,2)}}\sum_{\mathbf{\ell }}(-1)^{\mathbf{\ell }_{1}\mathbf{\ell }%
_{2}}\exp (i\mathbf{P}^{(\mathbf{p},r,\sigma )}.\mathbf{\ell )\ }T_{\mathbf{%
\ell }}\times   \label{indep} \\
&&\left\{ \sum_{u=0,1}\sum_{v=-1,0,1}\exp (\frac{5\pi i}{6}uv+i\mathbf{P}^{(%
\mathbf{p},r,\sigma )}.(u\mathbf{c}_{1}+v\mathbf{c}_{2})\mathbf{)}T_{u%
\mathbf{c}_{1}+v\mathbf{c}_{2}}\phi (\mathbf{x})\right\} \text{.}  \nonumber
\end{eqnarray}
Now, it can be noticed that the above expression differs form the definition
(\ref{fi}) only in that the ''seed'' function is changed by a superposition
of itself translated to the points $u\mathbf{c}_{1}+v\mathbf{c}_{2}.$ \
However, it was shown in Ref. (16) that the basis functions are completely
independent of an arbitrary change in the ''seed'' for any other function in
the first Landau level.

An alternative proof of this curious property is possible.\ To see it let us
represent the ''seed'' functions in (\ref{indep}) by their $\varphi _{%
\mathbf{p}}$-transform (\ref{transform}), yielding
\begin{eqnarray*}
\chi _{\mathbf{p}}^{(r,\sigma )}(x) &=&\frac{1}{\sqrt{6}N_{\mathbf{p}%
}^{(3,2)}}\sum_{\alpha =0,1}\sum_{\beta =-1,0,1}\exp (\frac{5\pi i}{6}uv+i%
\mathbf{P}^{(\mathbf{p},r,\sigma )}.(uc_{1}+vc_{2})\mathbf{)\times } \\
&&\int \frac{d\mathbf{q}}{(2\pi )^{2}}\sqrt{2\pi r_{0}^{2}}\bar{\varphi}_{%
\mathbf{q}}^{*}(u\mathbf{c}_{1}+v\mathbf{c}_{2})\bar{\varphi}_{\mathbf{q}}(%
\mathbf{x})\sum_{\mathbf{\ell }}\exp (i(\mathbf{P}^{(\mathbf{p},r,\sigma )}-%
\mathbf{q}).\mathbf{\ell )}\text{ ,}
\end{eqnarray*}
after employing the character of eigenfunctions of the operators $T_{\mathbf{%
\ell }}$ that the functions $\bar{\varphi}_{\mathbf{q}}$ have. \ But now the
following identity can be employed

\[
\sum_{\mathbf{\ell \in A}}\exp (i\mathbf{((q-q}^{\prime }\mathbf{).\ell )=}%
\frac{(2\pi )^{2}}{2\pi r_{0}^{2}}\delta (\mathbf{q,q}^{\prime }),
\]
where the $\delta $ function is non-vanishing for equal arguments modulo a
vector of the reciprocal lattice $Q=\frac{3}{2}Q_{1}\mathbf{t}_{1}+\frac{2}{3%
}Q_{2}\mathbf{t}_{2}.$ This relation allows to show the desired connection
between the basis functions $\chi _{\mathbf{p}}^{(r,\sigma )}$and $\bar{%
\varphi}_{\mathbf{p}}(\mathbf{x})$,

\[
\chi _{\mathbf{p}}^{(r,\sigma )}(x)=\mathcal{F}(\mathbf{p},r,\sigma )\ \bar{%
\varphi}_{\mathbf{P}^{(\mathbf{p},r,\sigma )}}(\mathbf{x}),
\]
where the phase factor is given by
\[
\mathcal{F}(\mathbf{p},r,\sigma )=\frac{1}{\sqrt{6}\sqrt{2\pi r_{0}^{2}}N_{%
\mathbf{p}}^{(3,2)}}\sum_{u=0,1}\sum_{v=-1,0,1}\exp (\frac{5\pi i}{6}uv+i%
\mathbf{P}^{(\mathbf{p},r,\sigma )}.(u\mathbf{c}_{1}+v\mathbf{c}_{2})\mathbf{%
)\ }\bar{\varphi}_{\mathbf{P}^{(\mathbf{p},r,\sigma )}}^{*}(u\mathbf{c}_{1}+v%
\mathbf{c}_{2})\text{ .}
\]

\bigskip

\subsection{\protect\bigskip Periodicity of the Slater determinants of
triplet orbitals}

Here we show that the single particle density associated with a Slater
determinant formed with a set of functions, each one corresponding to an
arbitrary linear combinations of the triplet of orbitals characterized by
the indices $(\mathbf{p},\sigma ),$ is periodic over the lattice $\mathbf{R.}
$ To consider this question, let us use the basic property that the density
of a many electron Slater determinant constructed with orthogonal orbitals
is the sum of the individual densities of each orbital. Writing the selected
linear combinations as
\begin{equation}
\Psi _{\mathbf{p}}^{(\sigma )}(\mathbf{x})=\sum_{r}C_{\mathbf{p}}^{(r,\sigma
)}\chi _{\mathbf{p}}^{(r,\sigma )}(\mathbf{x}),
\end{equation}
where the coefficients $C_{\mathbf{p}}^{(r,\sigma )}$ are arbitary, and
considering the symmetry properties (\ref{symmetry}), it follows that

\begin{eqnarray*}
\rho (\mathbf{x}) &=&\sum_{\mathbf{p}}\sum_{\sigma =\pm 1}|\Psi _{\mathbf{p}%
}^{(\sigma )}(\mathbf{x})|^{2} \\
&=&\sum_{\mathbf{p}}\sum_{\sigma =\pm 1}|\Psi _{\mathbf{p}}^{(-\sigma )}(%
\mathbf{x})|^{2} \\
&=&\sum_{\mathbf{p}}\sum_{\sigma =\pm 1}\sum_{r}\sum_{r^{\prime }}C_{\mathbf{%
p}}^{(r^{\prime },\sigma )*}C_{\mathbf{p}}^{(r,\sigma )}\left( \exp (i%
\mathbf{p.a}_{1})T_{a_{1}}\chi _{\mathbf{p}}^{(r^{\prime },\sigma )}(\mathbf{%
x})\right) ^{*}\exp (i\mathbf{p.a}_{1})T_{a_{1}}\chi _{\mathbf{p}%
}^{(r,\sigma )}(\mathbf{x}) \\
&=&\sum_{\mathbf{p}}\sum_{\sigma =\pm 1}\sum_{r}\sum_{r^{\prime }}C_{\mathbf{%
p}}^{(r^{\prime },\sigma )*}C_{\mathbf{p}}^{(r,\sigma )}\left( \exp (\frac{ie%
}{\hbar c}\mathbf{A(a}_{1}\mathbf{).x})\chi _{\mathbf{p}}^{(r^{\prime
},\sigma )}(\mathbf{x-a}_{1})\right) ^{*}\exp (\frac{ie}{\hbar c}\mathbf{A(a}%
_{1}\mathbf{).x})\chi _{\mathbf{p}}^{(r,\sigma )}(\mathbf{x-a}_{1}) \\
&=&\sum_{\mathbf{p}}\sum_{\sigma =\pm 1}|\Psi _{\mathbf{p}}^{(\sigma )}(%
\mathbf{x-a}_{1})|^{2} \\
&=&\rho (\mathbf{x-a}_{1}).
\end{eqnarray*}
One can show in a similar way that $\rho (\mathbf{x})=\rho (\mathbf{x-a}%
_{2}).$

\section{Appendix B}

Here we shall derive the general formula expressing the single particle
Hartree-Fock hamiltonian as a sum of\ magnetic translation operations. The
arguments of these operators are spatial vectors determined by the
reciprocal lattice associated with the Fourier components of the density. It
is worth remarking here that the lattice considered in this Appendix is not
restricted in any way. Therefore, the representation of the single particle
HF hamiltonian is valid for a general though periodic mean field problem.

Consider the action of the kernels of the direct and exchange interactions
defining the HF hamiltonian\ over any of the elements $\varphi _{p}$ of the
complete basis functions as follows

\begin{eqnarray}
\stackrel{\sim }{\varphi }^{(d)}(\mathbf{x}) &=&H_{HF}^{(d)}\ \varphi _{%
\mathbf{p}}(\mathbf{y}^{\prime })=\frac{e^{2}}{\varepsilon _{o}}\int d%
\mathbf{y}\ d\mathbf{y}^{\prime }\mathbf{\ }P(\mathbf{x},\mathbf{y}^{\prime
})\frac{\rho (\mathbf{y})-n_{0}/e^{2}}{\mid \mathbf{y}-\mathbf{y}^{\prime
}\mid }\varphi _{\mathbf{p}}(\mathbf{y}^{\prime }),  \label{direct} \\
\stackrel{\sim }{\varphi }^{(e)}(\mathbf{x}) &=&H_{HF}^{(e)}\ \varphi _{%
\mathbf{p}}(\mathbf{y}^{\prime })=-\frac{e^{2}}{\varepsilon _{o}}\int d%
\mathbf{y}\ d\mathbf{y}^{\prime }\mathbf{\ }P(\mathbf{x},\mathbf{y}^{\prime
})\frac{\rho (\mathbf{y,y}^{\prime })}{\mid \mathbf{y}-\mathbf{y}^{\prime
}\mid }\varphi _{\mathbf{p}}(\mathbf{y}^{\prime }),  \label{exchange}
\end{eqnarray}
where $n_{0}$ is the jellium background charge density making the overall
system neutral. The representation we shall derive is valid for both direct
and exchange kernels and will be discussed below for each case separately.

\subsection{Direct Coulomb interaction}

The potential term representing the direct Coulomb interaction can be
written in its Fourier transform representation as follows

\begin{eqnarray}
v^{(d)}(\mathbf{x}) &=&\frac{e^{2}}{\varepsilon _{o}}\int \ d\mathbf{x}%
^{\prime }\mathbf{\ }\frac{\rho (\mathbf{x}^{\prime })-n_{0}/e^{2}}{\mid
\mathbf{x}-\mathbf{x}^{\prime }\mid }  \nonumber \\
&=&\frac{e^{2}}{\varepsilon _{o}}\sum_{\mathbf{Q}\neq 0}\rho (\mathbf{Q}%
)\int \ d\mathbf{x}^{\prime }\mathbf{\ }\frac{\exp (i\mathbf{Q.x)}}{\mid
\mathbf{x}-\mathbf{x}^{\prime }\mid }  \nonumber \\
&=&\sum_{\mathbf{Q}}\frac{2\pi e^{2}}{\varepsilon _{o}}\frac{\rho (\mathbf{Q}%
)(1-\delta _{\mathbf{Q,0}})}{\mid \mathbf{Q}\mid }\exp (i\mathbf{Q.x)}
\nonumber \\
&=&\sum_{\mathbf{Q}}v^{(d)}(\mathbf{Q})\exp (i\mathbf{Q.x).}  \label{dirpot}
\end{eqnarray}
Then, using the following representation of the kernel for the projection
operator in the first Landau level

\[
P(\mathbf{x},\mathbf{x}^{\prime })=\frac{1}{2\pi r_{0}^{2}}\exp (-\frac{(%
\mathbf{x-x}^{\prime })^{2}}{4r_{0}^{2}})\exp (\frac{ie}{\hbar c}\mathbf{A(x}%
^{\prime }\mathbf{).x})
\]
we obtain,
\begin{equation}
\stackrel{\sim }{\varphi _{p}}^{(d)}(\mathbf{x})=\int \ d\mathbf{y}^{\prime }%
\mathbf{\ }P(\mathbf{x},\mathbf{y}^{\prime })\sum_{\mathbf{Q}}v^{(d)}(%
\mathbf{Q})\exp (i\mathbf{Q.y}^{\prime }\mathbf{)}\varphi _{\mathbf{p}}(%
\mathbf{y}^{\prime }).  \label{direct1}
\end{equation}
Recalling definition (\ref{fi}) we obtain for each translation term
\begin{eqnarray}
L_{\mathbf{\ell }}^{(d)}(\mathbf{x}) &=&\int d\mathbf{y}^{\prime }\mathbf{\ }%
P(\mathbf{x},\mathbf{y}^{\prime })\sum_{\mathbf{Q}}v^{(d)}(\mathbf{Q})\exp (i%
\mathbf{Q.y}^{\prime }\mathbf{)}T_{\mathbf{\ell }}\ \phi (\mathbf{y}^{\prime
}),  \nonumber \\
&=&\frac{1}{\sqrt{2\pi }r_{0}}\sum_{\mathbf{Q}}v^{(d)}(\mathbf{Q})\exp (-%
\frac{1}{4}r_{0}^{2}\mathbf{Q}^{2}\mathbf{)}\exp {\Large (}i\frac{\mathbf{%
Q.\ell }}{2}{\Large )}\exp {\Large (}-\frac{1}{4r_{0}^{2}}(\mathbf{x-(\ell
-r_{0}^{2}Q\times n)})^{2}{\Large )},  \nonumber \\
&&-\frac{i}{2r_{0}^{2}}\mathbf{x.(\ell +r_{0}^{2}n\times Q)}),  \nonumber \\
&=&\sum_{\mathbf{Q}}v^{(d)}(\mathbf{Q})\exp (-\frac{1}{4}r_{0}^{2}\mathbf{Q}%
^{2}\mathbf{)}\exp (i\frac{\mathbf{Q.l}}{2}\mathbf{)}T_{\mathbf{\ell
+r_{0}^{2}n\times Q}}\phi (\mathbf{x}).
\end{eqnarray}
This last relation can be further simplified by employing relation (\ref
{transl1}) and the similar property
\[
T_{\mathbf{R}_{1}}T_{\mathbf{R}_{2}}=\exp (\frac{ie}{\hbar c}\mathbf{%
A(R_{1}).R_{2})}T_{\mathbf{R}_{1}+\mathbf{R}_{2}},
\]
leading to the formula

\[
\int d\mathbf{y}^{\prime }\mathbf{\ }P(\mathbf{x},\mathbf{y}^{\prime })\sum_{%
\mathbf{Q}}v^{(d)}(\mathbf{Q})\exp (i\mathbf{Q.y}^{\prime }\mathbf{)}T_{%
\mathbf{\ell }}\phi (\mathbf{x})=\sum_{\mathbf{Q}}v^{(d)}(\mathbf{Q})\exp (-%
\frac{1}{4}r_{0}^{2}\mathbf{Q}^{2}\mathbf{)}T_{\mathbf{r_{0}^{2}n\times Q}%
}T_{\mathbf{\ell }}\phi (\mathbf{x}).
\]
After adding corresponding the integrals over all translations $\mathbf{\ell
}$ one obtains \

\[
H_{HF}^{(d)}\ \varphi _{\mathbf{p}}(\mathbf{x})=\sum_{\mathbf{Q}}v^{(d)}(%
\mathbf{Q})\exp (-\frac{1}{4}r_{0}^{2}\mathbf{Q}^{2}\mathbf{)}T_{\mathbf{%
r_{0}^{2}n\times Q}}\varphi _{\mathbf{p}}(\mathbf{x}).
\]
Since the equality is valid for any element of the complete basis $\varphi _{%
\mathbf{p}}(\mathbf{x})$ it then follows,
\begin{equation}
H_{HF}^{(d)}\ =\sum_{\mathbf{Q}}v^{(d)}(\mathbf{Q})\exp (-\frac{1}{4}%
r_{0}^{2}\mathbf{Q}^{2}\mathbf{)}T_{\mathbf{r_{0}^{2}\ n\times Q}}.
\label{directterm}
\end{equation}

\subsection{Exchange interaction}

\bigskip

The derivation of the analogous representation for the exchange interaction
kernel is more involved and needs for some special properties of the one
particle density matrix. Then, let us initially consider these properties.

\subsubsection{ One particle density matrix transformations}

\

The definition of the one-particle density matrix in terms of the Slater
determinant $\Phi (\mathbf{x}_{1}\mathbf{,x}_{2}\mathbf{,x}_{3}\mathbf{,x}%
_{4}\mathbf{,..x}_{N})$ can be transformed by performing a simultaneous
magnetic translation operation in a vector $\mathbf{R}$ of the periodic
lattice over all the particle coordinates. Since this is assumed to be a
symmetry transformation of the system, this map should leave the many
particle state invariant. Thus by assumption

\begin{eqnarray}
\rho (\mathbf{x}_{1}\mathbf{,x}_{2}) &=&\int ...\int d\mathbf{x}_{3}d\mathbf{%
x}_{3}..d\mathbf{x}_{N}\ |\Phi (\mathbf{x}_{1}\mathbf{,x}_{2}\mathbf{,x}_{3}%
\mathbf{,x}_{4}\mathbf{,..x}_{N})|^{2} \\
&=&\int ...\int d\mathbf{x}_{3}d\mathbf{x}_{3}..d\mathbf{x}_{N}\
|\prod_{i=3}^{N}T_{R}(\mathbf{x}_{i})\Phi (\mathbf{x}_{1}\mathbf{,x}_{2}%
\mathbf{,x}_{3}\mathbf{,x}_{4}\mathbf{,..x}_{N})|^{2}  \nonumber \\
&=&\sum_{\varkappa }\Psi _{\varkappa }(\mathbf{x}_{1})\Psi _{\varkappa }^{*}(%
\mathbf{x}_{2}),  \nonumber \\
&=&\sum_{\varkappa }T_{R}(\mathbf{x}_{1})\Psi _{\varkappa }(\mathbf{x}%
_{1})\left( T_{R}(\mathbf{x}_{2})\Psi _{\varkappa }(\mathbf{x}_{2})\right)
^{*},  \nonumber \\
&=&\sum_{\varkappa }\exp (\frac{ie}{\hbar c}\mathbf{A(R).(x}_{1}-\mathbf{x}%
_{2}))\exp (\Psi _{\varkappa }(\mathbf{x}_{1}-R)\left( \Psi _{\varkappa }(%
\mathbf{x}_{2}-R)\right) ^{*}\rho (\mathbf{x}_{1}\mathbf{,x}_{2}),  \nonumber
\\
&=&\exp (\frac{ie}{\hbar c}\mathbf{A(R).(x}_{1}-\mathbf{x}_{2}))\rho (%
\mathbf{x}_{1}-R\mathbf{,x}_{2}-R),  \nonumber
\end{eqnarray}
an expression that furnishes the transformation law of the one particle
density matrix under spatial shifts in the vectors of the periodic lattice.

\subsubsection{ The density determines the density matrix}

Below we will show that the Fourier components of the density completely
determine the whole one-particle density matrix. \ For this purpose, let us
use the new variable $\mathbf{z=x-x}^{\prime }$ in the one-particle density
matrix, so that $\rho (\mathbf{x,x}^{\prime })=\stackrel{\sim }{\rho }(%
\mathbf{x,z}),$ where the tilde over $\rho $ underlines the different
functional expression of the new definition. It now transforms under spatial
shifts as
\begin{equation}
\ \stackrel{\sim }{\rho }(\mathbf{x-R,z})=\exp (-\frac{ie}{\hbar c}\mathbf{%
A(R).z})\stackrel{\sim }{\rho }(\mathbf{x,z}).
\end{equation}
Therefore, the function

\[
\stackrel{\sim }{g}(\mathbf{x,z})=\exp (-\frac{ie}{\hbar c}\mathbf{A(x).z})%
\stackrel{\sim }{\rho }(\mathbf{x,z}),
\]
is fully periodic in the variable $\mathbf{x}$ under the lattice shifts.
That is $\stackrel{\sim }{g}(\mathbf{x-R,z})=\stackrel{\sim }{g}(\mathbf{x,z}%
).$ Then, Fourier expanding $\stackrel{\sim }{g}$ \ leads to the following
expression for $\stackrel{\sim }{\rho }$

\[
\stackrel{\sim }{\rho }(\mathbf{x,z})=\sum_{\mathbf{Q}}\exp (i(\mathbf{Q}-%
\frac{e}{\hbar c}\mathbf{A(R)).x})\stackrel{\sim }{\rho }(\mathbf{Q,z}),
\]
where

\begin{equation}
\stackrel{\sim }{\rho }(\mathbf{Q,z})=\frac{1}{A_{cell}}\int d\mathbf{x}%
^{\prime }\exp (-i\mathbf{Q.x}^{\prime }\mathbf{)}\exp (i\frac{e}{\hbar c}%
\mathbf{A(z).x}^{\prime })\stackrel{\sim }{\rho }(\mathbf{x}^{\prime }%
\mathbf{,z}).  \label{elele}
\end{equation}
Further, let us consider the density written as the sum
\[
\sum_{\varkappa }\Psi _{\varkappa }(\mathbf{x})\Psi _{\varkappa }^{*}(%
\mathbf{x}^{\prime })=\sum_{\mathbf{p}}\sum_{r,\sigma ,r^{\prime },\sigma
^{\prime }}f_{r,\sigma }(\mathbf{p)}\chi _{\mathbf{p}}^{(r,\sigma )}(\mathbf{%
x}\Bbb{{)}f_{r^{\prime },\sigma ^{\prime }}^{*}(\mathbf{p)}\chi _{\mathbf{p}%
}^{(r^{\prime },\sigma )}(\mathbf{x}^{\prime })^{*},}
\]
and the fact that each $\chi _{\mathbf{p}}^{(r^{\prime },\sigma )}$ in the
lower Landau level can be expanded as a linear combination of the complete
basis functions$\ \{T_{\mathbf{\ell }}\ \phi (\mathbf{x})\}$ defined over a
lattice with one flux quantum per unit cell.\cite{tao} Then, a generic term
in the integral in (\ref{elele} ) has the form
\[
\stackrel{\sim }{\rho _{\mathbf{\ell ,\ell }^{\prime }}}(\mathbf{Q,z})=\frac{%
1}{A_{cell}}\int d\mathbf{x}\exp (-i\mathbf{Q.x)}\exp (i\frac{e}{\hbar c}%
\mathbf{A(z).x})\mathbf{\ }T_{\mathbf{\ell }}\ \phi (\mathbf{x})(T_{\mathbf{%
\ell }^{\prime }}\phi (\mathbf{x}^{\prime })\mid _{\mathbf{x}^{\prime }=%
\mathbf{x-z}})^{*}.
\]
Since the function $\phi $ is a Gaussian, its magnetic translations (\ref
{magntra}) have also this character, and thus the integrals in (\ref{elele})
can be calculated explicitly. We obtain,
\[
\stackrel{\sim }{\rho _{\mathbf{\ell ,\ell }^{\prime }}}(\mathbf{Q,z})=%
\stackrel{\sim }{\rho _{\mathbf{\ell ,\ell }^{\prime }}}(\mathbf{Q,0})\exp (-%
\frac{\mathbf{z}^{2}}{4r_{0}^{2}}-\frac{1}{4r_{0}^{2}}\mathbf{Q.(n\times z+}%
i\ \mathbf{z})).
\]
But since the phase factor here is independent of the indices $\mathbf{\ell
,\ell }^{\prime }$ ,\ the summation of all the contributions in (\ref{elele}%
) leads to the relation

\begin{eqnarray}
\stackrel{\sim }{\rho }(\mathbf{Q,z}) &=&\rho (\mathbf{Q})\exp (-\frac{%
\mathbf{z}^{2}}{4r_{0}^{2}}-\frac{1}{4r_{0}^{2}}\mathbf{Q.(n\times z+}i\
\mathbf{z})),  \label{onepart} \\
\stackrel{\sim }{\rho }(\mathbf{Q,0}) &=&\rho (\mathbf{Q}),  \nonumber
\end{eqnarray}
which expresses the interesting result that the full one-particle density
matrix is completely determined by the particle density.

We now consider the exchange term (\ref{exchange}). Using the above results
the integrals in

\begin{equation}
\stackrel{\sim }{\varphi }^{(e)}(\mathbf{x})=\ -e^{2}\int d\mathbf{y}%
^{\prime }\mathbf{\ }P(\mathbf{x},\mathbf{y}^{\prime })\int d\mathbf{z}\sum_{%
\mathbf{Q}}\exp (i(\mathbf{Q}-\frac{e}{\hbar c}\mathbf{A(z)).y}^{\prime })%
\frac{\stackrel{\sim }{\rho }(\mathbf{Q,z})}{\mid \mathbf{z}\mid }\varphi _{%
\mathbf{p}}(\mathbf{y}^{\prime }-\mathbf{z}),
\end{equation}
can be calculated\ explicitly. For this purpose, let us consider again the
integral of a generic term $T_{\mathbf{\ell }}\ \phi $ in the sum over $%
\mathbf{\ell }$ defining the functions $\varphi _{\mathbf{p}}$. The integral
for this term is then reduced to two simple integrals, which after some
algebra can be explicitly evaluated to obtain
\begin{eqnarray}
L^{(e)}(\mathbf{x}) &=&\sum_{\mathbf{Q}}v^{(e)}(\mathbf{Q})\exp (-\frac{1}{4}%
r_{0}^{2}\mathbf{Q}^{2}\mathbf{)}\exp (i\frac{\mathbf{Q.\ell }}{2}\mathbf{)}%
T_{\mathbf{\ell +r_{0}^{2}n\times Q}}\phi (\mathbf{x}),  \nonumber \\
&=&\sum_{\mathbf{Q}}v^{(e)}(\mathbf{Q})\exp (-\frac{1}{4}r_{0}^{2}\mathbf{Q}%
^{2}\mathbf{)}T_{\mathbf{r_{0}^{2}n\times Q}}T_{\mathbf{\ell }}\phi (\mathbf{%
x}), \\
v^{(e)}(\mathbf{Q}) &=&-\frac{2\pi e^{2}r_{0}}{\varepsilon _{o}}\sqrt{\frac{%
\pi }{2}}\stackrel{\sim }{\rho }(\mathbf{Q,0})\ \exp (\frac{1}{4}r_{0}^{2}%
\mathbf{Q}^{2}\mathbf{)\ }I_{0}(\frac{1}{4}r_{0}^{2}\mathbf{Q}^{2}).
\end{eqnarray}
Henceforth, after adding all terms for different values of $\mathbf{\ell }$
follows
\[
H_{HF}^{(e)}\varphi _{p}(\mathbf{x})=\left( \sum_{\mathbf{Q}}v^{(e)}(\mathbf{%
Q})\exp (-\frac{1}{4}r_{0}^{2}\mathbf{Q}^{2}\mathbf{)}T_{\mathbf{%
r_{0}^{2}n\times Q}}\right) \varphi _{p}(\mathbf{x}).
\]
Using the independence and completeness of the basis formed by the $\varphi
_{p}$'s in the first Landau level, we thus obtain
\[
H_{HF}^{(e)}=\sum_{\mathbf{Q}}v^{(e)}(\mathbf{Q})\exp (-\frac{1}{4}r_{0}^{2}%
\mathbf{Q}^{2}\mathbf{)}T_{\mathbf{r_{0}^{2}n\times Q}}.
\]
Combining this results and the previous one (\ref{directterm}) the following
representation for the one-particle Hartree-Fock Hamiltonian in the first
Landau level follows,

\begin{eqnarray}
H_{HF} &=&\sum_{\mathbf{Q}}(v^{(d)}(\mathbf{Q})+v^{(e)}(\mathbf{Q}))\exp (-%
\frac{1}{4}r_{0}^{2}\mathbf{Q}^{2}\mathbf{)}T_{\mathbf{r_{0}^{2}n\times Q}}
\nonumber \\
&=&\sum_{\mathbf{Q}}v(\mathbf{Q})\exp (-\frac{1}{4}r_{0}^{2}\mathbf{Q}^{2}%
\mathbf{)}T_{\mathbf{r_{0}^{2}n\times Q}},  \label{ham}
\end{eqnarray}
with the coefficients defined by

\begin{equation}
v(\mathbf{Q})=2\pi r_{0}^{2}\rho (\mathbf{Q})\left( \frac{(1-\delta _{%
\mathbf{Q,0}})}{r_{0}\mid \mathbf{Q}\mid }-\sqrt{\frac{\pi }{2}}\ \exp (%
\frac{1}{4}r_{0}^{2}\mathbf{Q}^{2}\mathbf{)\ }I_{0}(\frac{1}{4}r_{0}^{2}%
\mathbf{Q}^{2})\right) \frac{e^{2}}{\varepsilon _{o}r_{0}}.
\end{equation}

\section{Appendix C}

In this section we prove that the eigenfunctions that vanish at all latice
points $\mathbf{R}$ are solutions of the HF equations$\mathbf{.}$ For this
purpose, let us consider the Fourier components
\[
\rho _{\mathbf{y}}(\mathbf{Q})=\frac{1}{A_{cell}}\int d\mathbf{x\ }\rho _{%
\mathbf{y}}(\mathbf{x})\exp (-i\mathbf{Q.x})
\]
of the configuration space periodically extended density operator

\begin{eqnarray*}
\rho _{\mathbf{y}}(\mathbf{x}) &=&\delta (\mathbf{x-y})=\sum_{\mathbf{R}%
}\delta (\mathbf{x-y+R}) \\
&=&\frac{1}{A_{cell}}\sum_{\mathbf{Q}}\exp (-i\mathbf{Q.y})\exp (i\mathbf{Q.x%
}).
\end{eqnarray*}
We will examine the projection of this operator in the first Landau level.
For this aim it is useful to consider the set of operators
\begin{eqnarray*}
P_{i} &=&\partial _{i}+i\frac{e}{\hbar c}A_{i}\mathbf{(x)} \\
&=&\partial _{i}+\frac{i}{2r_{0}^{2}}\epsilon _{ij}x_{j} \\
G_{i} &=&\partial _{i}-i\frac{e}{\hbar c}A_{i}\mathbf{(x)} \\
&=&\partial _{i}-\frac{i}{2r_{0}^{2}}\epsilon _{ij}x_{j} \\
\epsilon _{ij} &\equiv &\left(
\begin{tabular}{ll}
0 & -1 \\
1 & 0
\end{tabular}
\right) ,
\end{eqnarray*}
where the indices $i,j$\ take the values 1,2. These quantities satisfy the
commutation relations

\begin{eqnarray*}
\left[ P_{i},G_{j}\right] &=&0, \\
\left[ P_{i},P_{j}\right] &=&-\frac{\epsilon _{ij}}{r_{0}^{2}}, \\
\left[ G_{i},G_{j}\right] &=&\frac{\epsilon _{ij}}{r_{0}^{2}}.
\end{eqnarray*}
With their use, the spatial Fourier modes can be identically represented as

\begin{eqnarray}
\exp (-i\mathbf{Q.x}) &=&\exp (-\frac{1}{2}(P_{i}-G_{i}).\epsilon _{ik}Q_{k}%
\mathbf{(}2r_{0}^{2}))  \nonumber \\
&=&\exp (-\frac{1}{2}(\mathbf{P-G}).\mathbf{n}\times \mathbf{Q(}2r_{0}^{2})),
\label{project}
\end{eqnarray}
where the 3D-vectors $\mathbf{P}$ and $\mathbf{G}$ are defined as $%
(P_{1},P_{2},0)$ and $(G_{1},G_{2},0)$. But the two components $P_{i}$ are
linear combinations of the rising and lowering operators which transform
functions of a Landau level in functions of the higher and lower contiguous
levels. Therefore, the projection onto the first Landau level of the Fourier
component can be obtained by simply omitting the operator $\mathbf{P}$ in
the above formula. Then, the projection of the density operator will be
\begin{eqnarray*}
{\rho }_{\mathbf{y}}^{(p)}(\mathbf{x}) &=&\frac{1}{A_{cell}}\sum_{\mathbf{Q}%
}\exp (i\mathbf{Q.y})\exp (-\frac{1}{2}\mathbf{G}.\mathbf{n}\times \mathbf{Q(%
}2r_{0}^{2})) \\
&=&\frac{1}{A_{cell}}\sum_{\mathbf{Q}}\exp (i\mathbf{Q.y})T_{\mathbf{n}%
\times \mathbf{Q(}r_{0}^{2})} \\
&=&\frac{1}{A_{cell}}\sum_{\mathbf{x}^{*}}\exp (i\mathbf{x}^{*}\mathbf{.}%
\frac{\mathbf{n\times y}}{r_{0}^{2}})T_{\mathbf{x}^{*}},
\end{eqnarray*}
where a new set of spatial lattice vectors was defined in terms of the
Fourier wavevectors,
\[
\mathbf{x}^{*}\mathbf{=n}\times \mathbf{Q}r_{0}^{2}.
\]
Let us consider now the identity
\[
\rho _{\mathbf{0}}\sum_{r}g_{r}^{0}(\mathbf{p)}\mid \!\!\chi _{\mathbf{p}%
}^{(r,\sigma )}\rangle =0,
\]
where the three coefficients $g_{r}^{0}(\mathbf{p)},\ r=1,2,3,$ are fixed by
the condition that the linear combination of the functions of the two
triplets of \ functions $\mid \!\!\chi _{\mathbf{p}}^{(r,\sigma )}\rangle \ $%
for $\sigma =\pm 1$ vanish at the origin $\mathbf{x}=\mathbf{0}.$ Projecting
the above identity onto the first Landau level by acting with $P$ at the
left it follows
\begin{eqnarray}
P\rho _{\mathbf{0}}\sum_{r}g_{r}^{0}(\mathbf{p)} &\mid &\!\!\chi _{\mathbf{p}%
}^{(r,\sigma )}\rangle =P\rho _{\mathbf{0}}PP\sum_{r}g_{r}^{0}(\mathbf{p)}%
\mid \chi _{\mathbf{p}}^{(r,\sigma )}\rangle   \nonumber \\
&=&{\rho }_{\mathbf{0}}^{(p)}\sum_{r}g_{r}^{0}(\mathbf{p)}\mid \chi _{%
\mathbf{p}}^{(r,\sigma )}\rangle =0,  \label{vanish}
\end{eqnarray}
implying that the projected periodic density has the defined functions as
eigenvectors with zero eigenvalues. Let us inspect further the commutation
properties of the projected density operator. Using the representation of
the single particle Hamiltonian in terms of the magnetic translations found
in Appendix B and the relations
\begin{eqnarray*}
T_{\mathbf{x}^{*}}T_{\mathbf{y}^{*}} &=&\exp (\frac{ie}{\hbar c}\mathbf{%
A(x^{*}).y}^{*})T_{\mathbf{x}^{*}+\mathbf{y}^{*}}, \\
T_{\mathbf{y}^{*}}T_{\mathbf{x}^{*}} &=&\exp (-\frac{ie}{\hbar c}\mathbf{%
A(x^{*}).y}^{*})T_{\mathbf{x}^{*}+\mathbf{y}^{*}},
\end{eqnarray*}
it follows that

\begin{eqnarray}
\left[ H_{HF},{\rho }_{\mathbf{0}}^{(p)}\right]  &=&\sum_{\mathbf{x}%
^{*}}\sum_{\mathbf{y}^{*}}u\text{(}\mathbf{x}^{*})\left[ T_{\mathbf{x}%
^{*}},T_{\mathbf{y}^{*}}\right]  \\
&=&\sum_{\mathbf{x}^{*}}\sum_{\mathbf{z}^{*}}u\text{(}\mathbf{x}^{*})(\exp (%
\frac{ie}{\hbar c}\mathbf{A(x^{*}).z}^{*})-\exp (-\frac{ie}{\hbar c}\mathbf{%
A(x^{*}).z}^{*}))T_{\mathbf{z}^{*}}.  \nonumber
\end{eqnarray}
\ Considering that $u$ is the coefficient of the superposition of
translations giving the single particle HF Hamiltonian which is a function
of $\mathbf{Q}^{2}$, and using the inversion symmetry of the solution under
consideration $u$($\mathbf{x}^{*})=u$(-$\mathbf{x}^{*}),$ one finds
\begin{equation}
\left[ H_{HF},{\rho }_{\mathbf{0}}^{(p)}\right] =0.
\end{equation}
Hence, acting with the Hamiltonian on (\ref{vanish}) and employing the
definition of the projection operators and the periodic density operator it
follows that

\begin{eqnarray*}
H_{HF}{\rho }^{(p)}_{\mathbf{0}}\sum_{r}g_{r}^{0}(\mathbf{p)} &\mid &\chi _{%
\mathbf{p}}^{(r,\sigma )}\rangle =\tilde{\rho}_{\mathbf{0}%
}H_{HF}\sum_{r}g_{r}^{0}(\mathbf{p)}\mid \chi _{\mathbf{p}}^{(r,\sigma
)}\rangle \\
&\equiv &\int dx^{\prime }P(\mathbf{x,x}^{\prime })\rho _{\mathbf{0}}(%
\mathbf{x}^{\prime })\sum_{r}h_{r}(\mathbf{p})\chi _{\mathbf{p}}^{(r,\sigma
)}(\mathbf{x}^{\prime }) \\
&=&\int dx^{\prime }P(\mathbf{x,x}^{\prime })\sum_{R}\delta (\mathbf{x}%
^{\prime }\mathbf{-R})\sum_{r}h_{r}(\mathbf{p})\chi _{\mathbf{p}}^{(r,\sigma
)}(\mathbf{x}^{\prime }) \\
&=&\sum_{R}P(\mathbf{x,R})\sum_{r}h_{r}(\mathbf{p})\chi _{\mathbf{p}%
}^{(r,\sigma )}(\mathbf{R}) \\
&=&\sum_{R}\frac{1}{2\pi r_{0}^{2}}\exp (-\frac{(\mathbf{x-R})^{2}}{%
4r_{0}^{2}})\exp (\frac{ie}{\hbar c}\mathbf{A(R).x})\sum_{r}h_{r}(\mathbf{p}%
)\chi _{\mathbf{p}}^{(r,\sigma )}(\mathbf{R}) \\
&=&\sum_{R}\frac{1}{\sqrt{2\pi r_{0}^{2}}}\sum_{r}h_{r}(\mathbf{p})\chi _{%
\mathbf{p}}^{(r,\sigma )}(\mathbf{R})T_{\mathbf{R}}\phi (\mathbf{x}) \\
&=&0.
\end{eqnarray*}
Finally, taking into account that the set of functions $T_{\mathbf{R}}\phi
(x)$ for all values of the lattice (\ref{perlatt}) are linearly independent
\cite{tao}, it follows that

\begin{equation}
\sum_{r}h_{r}(\mathbf{p})\chi _{\mathbf{p}}^{(r,\sigma )}(\mathbf{R})=0.
\end{equation}
In addition, since the coefficients of the linear combinations of the
triplet functions for the two signs of $\sigma $ are uniquely defined by the
vanishing condition of the function at the origin (or any point of the
lattice $\mathbf{R}$), we have,
\[
h_{r}(\mathbf{p})=\varepsilon g_{r}^{0}(\mathbf{p)},
\]
where $\varepsilon $ is here a normalization constant. Finally, we obtain
\begin{eqnarray}
H_{HF}\sum_{r}g_{r}^{0}(\mathbf{p)} &\mid &\chi _{\mathbf{p}}^{(r,\sigma
)}\rangle =\sum_{r}h_{r}(p)\mid \chi _{\mathbf{p}}^{(r,\sigma )}\rangle , \\
&=&\varepsilon \sum_{r}g_{r}^{0}(p)\mid \chi _{\mathbf{p}}^{(r,\sigma
)}\rangle .  \nonumber
\end{eqnarray}
That is, the special functions that vanish at all latice points are
eigenfunctions of the one particle HF Hamiltonian.

\section{Appendix D}

In this Appendix we sketch the derivation of the formula (\ref{e2alter})
employed in the evaluation of the sample size dependence of the correction
to the energy per particle. As before, we use the shorthand notation
\begin{eqnarray*}
\Phi _{\alpha } &\equiv &\Phi _{\mathbf{p}_{\alpha }}^{(b_{\alpha },\sigma
_{\alpha })}=\sum_{r}g_{r}^{b_{\alpha }}(\mathbf{p}_{\alpha })\chi _{\mathbf{%
p}_{\alpha }}^{(r,\sigma _{\alpha })}(x) \\
&=&\sum_{r}d_{\mathbf{p}_{\alpha }}^{(b_{\alpha },r,\sigma _{\alpha })}\
\varphi _{\mathbf{p}_{\alpha },r}(\mathbf{x}), \\
d_{\mathbf{p}_{\alpha }}^{(b_{\alpha },r,\sigma _{\alpha })} &=&\mathcal{F}(%
\mathbf{p}_{\alpha },r,\sigma _{\alpha })\ g_{r}^{b_{\alpha }}(\mathbf{p}%
_{\alpha }) \\
\mathbf{P}_{\alpha ,r} &\equiv &\mathbf{P}^{(\mathbf{p}_{\alpha },r,\sigma
_{\alpha })}=\mathbf{p}_{\alpha }-r\ \mathbf{s}_{1}+\frac{\sigma _{\alpha }-1%
}{2}\frac{3}{2}\mathbf{s}_{2}.
\end{eqnarray*}
The relations expressing the functions $\chi _{\mathbf{p}_{\alpha
}}^{(r,\sigma _{\alpha })}$ in terms of $\varphi _{\mathbf{p}_{\alpha }}$%
times a phase factor has also been employed in order to define the new
coefficients $d_{\mathbf{p}_{\alpha }}^{(b_{\alpha },r,\sigma _{\alpha })}.$
\ By representing the Coulomb potential by its Fourier transform, and
substituting all the expansions of the functions $\varphi _{\mathbf{p}%
_{\alpha }}$ in terms of \ magnetic translations of the ''seed'' in the
matrix elements (\ref{matrix}), the resulting Gaussian integrals for each
term can be evaluated. After some algebra it is possible to obtain the
expresion

\begin{eqnarray}
M(\alpha ,\alpha ^{\prime } &\mid &\beta ^{\prime },\beta )=\int \int d%
\mathbf{x}d\mathbf{x}^{\prime }\,\,\Phi _{\alpha }^{*}(\mathbf{x})\Phi
_{\alpha ^{\prime }}^{*}(\mathbf{x}^{\prime })\,\frac{1}{\mid \mathbf{x}-%
\mathbf{x}^{\prime }\mid }\,\Phi _{\beta ^{\prime }}(\mathbf{x}^{\prime
})\Phi _{\beta }(\mathbf{x}),  \label{matrix1} \\
&=&\sum_{r_{\alpha }}\sum_{r_{\alpha ^{\prime }}}\sum_{r_{\beta ^{\prime
}}}\sum_{r_{\beta }}\left( d_{\mathbf{p}_{\alpha }}^{(b_{\alpha },r_{\alpha
},\sigma _{\alpha })}d_{\mathbf{p}_{\alpha ^{\prime }}}^{(b_{\alpha ^{\prime
}},r_{\alpha ^{\prime }},\sigma _{\alpha ^{\prime }})}\right) ^{*}d_{\mathbf{%
p}_{\beta ^{\prime }}}^{(0,r_{\beta ^{\prime }},\sigma _{\beta ^{\prime
}})}d_{\mathbf{p}_{\beta }}^{(0,r_{\beta },\sigma _{\beta })}\times
\nonumber \\
&&\int \int d\mathbf{x}d\mathbf{x}^{\prime }\,\,\varphi _{\mathbf{P}_{\alpha
},_{r_{\alpha }}}^{*}(\mathbf{x})\ \varphi _{\mathbf{P}_{\alpha ^{\prime
},r_{\alpha ^{\prime }}}}^{*}(\mathbf{x}^{\prime })\,\frac{1}{\mid \mathbf{x}%
-\mathbf{x}^{\prime }\mid }\varphi _{\mathbf{P}_{\beta ^{\prime },r_{\beta
^{\prime }}}}(\mathbf{x}^{\prime })\ \varphi _{\mathbf{P}_{\beta ,r_{\beta
}}}(\mathbf{x}),  \nonumber \\
&=&N_{\phi _{0}}\sum_{r_{\alpha }}\sum_{r_{\alpha ^{\prime }}}\sum_{r_{\beta
^{\prime }}}\sum_{r_{\beta }}\left( d_{\mathbf{p}_{\alpha }}^{(b_{\alpha
},r_{\alpha },\sigma _{\alpha })}d_{\mathbf{p}_{\alpha ^{\prime
}}}^{(b_{\alpha ^{\prime }},r_{\alpha ^{\prime }},\sigma _{\alpha ^{\prime
}})}\right) ^{*}d_{\mathbf{p}_{\beta ^{\prime }}}^{(0,r_{\beta ^{\prime
}},\sigma _{\beta ^{\prime }})}d_{\mathbf{p}_{\beta }}^{(0,r_{\beta },\sigma
_{\beta })}\times   \nonumber \\
&&\delta (\mathbf{P}_{\beta ^{\prime },r_{\beta ^{\prime }}}+\mathbf{P}%
_{\beta ,r_{\beta }}-\mathbf{P}_{a,r_{\alpha }}\mathbf{-P}_{\alpha ^{\prime
},r_{\alpha ^{\prime }}},0)\times   \nonumber \\
&&\frac{\varphi _{\mathbf{p}_{\alpha }}^{*}(\mathbf{n\times (P}_{\beta
^{\prime },r_{\beta ^{\prime }}}\mathbf{-P}_{\alpha ^{\prime },r_{\alpha
^{\prime }}}\mathbf{)}r_{0}^{2})\ \varphi _{\mathbf{p}_{\alpha ^{\prime
}}}^{*}(-\mathbf{n\times (P}_{\beta ^{\prime },r_{\beta ^{\prime }}}\mathbf{%
-P}_{\alpha ^{\prime },r_{\alpha ^{\prime }}}\mathbf{)}r_{0}^{2})}{N_{%
\mathbf{P}_{\beta ^{\prime },}r_{\beta ^{\prime }}}^{(2,3)}N_{\mathbf{P}%
_{\beta ,r_{\beta }}}^{(2,3)}}\,\times   \nonumber \\
&&\sum_{\mathbf{Q}^{*}}\frac{2\pi }{\mid \mathbf{P}_{\beta ^{\prime
},r_{\beta ^{\prime }}}\mathbf{-P}_{\alpha ^{\prime },r_{\alpha ^{\prime }}}+%
\mathbf{Q}^{*}\mid }\exp (-r_{0}^{2}(\mathbf{P}_{\beta ^{\prime },r_{\beta
^{\prime }}}\mathbf{-P}_{\alpha ^{\prime },r_{\alpha ^{\prime }}}+\mathbf{Q}%
^{*})^{2})\times \   \nonumber \\
&&\exp \left[ -i\ r_{0}^{2}\mathbf{n}\times \mathbf{Q}^{*}\cdot (\mathbf{P}%
_{\alpha ,r_{\alpha }}+\mathbf{P}_{\beta ,r_{\beta }}-\mathbf{P}_{\beta
^{\prime },r_{\beta ^{\prime }}}\mathbf{-P}_{\alpha ^{\prime },r_{\alpha
^{\prime }}})\right] ,  \nonumber
\end{eqnarray}
where the summation vectors $\mathbf{Q}^{*}$ and the norm functions are
defined as follows,
\begin{eqnarray}
\mathbf{Q}^{*} &=&n_{1}\frac{3}{2}\ \mathbf{s}_{1}+n_{2}\ \mathbf{s}_{2},%
\text{ \ } \\
N_{\mathbf{P}}^{(2,3)} &=&\sqrt{N_{\phi _{0}}}\sqrt{\sum_{\mathbf{\ell }%
}\left( -1\right) ^{\ell _{1}\ell _{2}}\exp (i\ \mathbf{P}.\mathbf{\ell -}%
\frac{\mathbf{\ell }^{2}}{4r_{0}^{2}}\mathbf{)\ }},  \nonumber \\
\mathbf{\ell } &=&\ \mathbf{\ell }_{1}(2\ \mathbf{c}_{1})+\mathbf{\ell }%
_{2}\ (3\ \mathbf{c}_{2}),  \nonumber
\end{eqnarray}
and, as before, $\delta (\mathbf{P}^{\prime },\mathbf{P})$ equals one when
the arguments are same modulo vectors of the class $\mathbf{Q}^{*}$ defined
above, and zero otherwise$.$ It is possible to show now that the ratio
multiplying this latter function a simply a phase factor. For this purpose
let us consider the definition of the $\varphi _{\mathbf{P}_{\alpha
,r_{\alpha }}},$

\begin{equation}
\varphi _{\mathbf{P}_{\alpha },r_{\alpha }}(\mathbf{x})=\frac{1}{N_{\mathbf{P%
}_{\alpha },r_{\alpha }}^{(2,3)}}\sum_{\mathbf{\ell }}\left( -1\right)
^{\ell _{1}\ell _{2}}\exp (i\ \mathbf{P}_{\alpha ,r_{\alpha }}.\mathbf{\ell
)\ }T\mathbf{_{{\ell }}\ \phi (x),}
\end{equation}
from which follows directly a connection between the value of this function
at the origin and the norm functions appearing in (\ref{matrix1})

\[
\varphi _{\mathbf{P}_{\alpha ,r_{\alpha }}}(\mathbf{0})=\sqrt{\sum_{\mathbf{%
\ell }}\left( -1\right) ^{\ell _{1}\ell _{2}}\exp (i\ \mathbf{P}_{\alpha
,r_{\alpha }}.\mathbf{\ell -}\frac{\mathbf{\ell }^{2}}{4r_{0}^{2}}\mathbf{)\
}}=\frac{N_{\mathbf{P}_{\alpha },_{r_{\alpha }}}^{(2,3)}}{\sqrt{2\pi
r_{0}^{2}}N_{\phi _{0}}}.
\]
Further, let us recall the formula connecting a translated function of the
basis $\varphi _{\mathbf{p}}$ and the same function for a shifted momenta (%
\ref{phase}). After considering this relation evaluated at $\mathbf{x}=0$,
the following expression can be obtained

\begin{eqnarray*}
\varphi _{\mathbf{P}_{\alpha ,r_{\alpha }}}(\mathbf{n\times (P}_{\beta
^{\prime },r_{\beta ^{\prime }}}\mathbf{-P}_{\alpha ^{\prime },r_{\alpha
^{\prime }}}\mathbf{)}r_{0}^{2}) &=&\varphi _{\mathbf{P}_{\alpha
},_{r_{\alpha }}}(\mathbf{n\times (P}_{\alpha ,r_{\alpha }}\mathbf{-P}%
_{_{\beta },_{r_{\beta }}}\mathbf{)}r_{0}^{2}) \\
&=&\mathcal{F}_{\mathbf{P}_{\alpha },_{r_{\alpha }}}(\mathbf{n\times (P}%
_{\beta ,r_{\beta }}\mathbf{-P}_{\alpha ,r_{\alpha }}\mathbf{)}%
r_{0}^{2})\varphi _{\mathbf{P_{\alpha ,r_{\alpha }}}-\frac{2e}{\hbar c}A(%
\mathbf{n\times (P_{\alpha ,r_{\alpha }}-P_{\beta ,r_{\beta }})}r_{0}^{2})}(%
\mathbf{0}).
\end{eqnarray*}
But the new momenta argument of the function in the rhs can be simplified as

\begin{eqnarray*}
-\frac{2e}{\hbar c}A(\mathbf{n\times (P}_{\alpha ,r_{\alpha }}\mathbf{-P}%
_{\beta ,r_{\beta }}\mathbf{)}r_{0}^{2}) &=&\frac{2}{r_{0}^{2}}\frac{\mathbf{%
n}}{2}\times (\mathbf{n\times (P}_{\alpha ,r_{\alpha }}\mathbf{-P}_{\beta
,r_{\beta }}\mathbf{)}r_{0}^{2}) \\
&=&\mathbf{P}_{\beta ,r_{\beta }}\mathbf{-P}_{a,r_{a}},
\end{eqnarray*}
to produce

\[
\varphi _{\mathbf{P}_{\alpha ,r_{\alpha }}}(\mathbf{n\times (P}_{\beta
^{\prime },r_{\beta ^{\prime }}}\mathbf{-P}_{\alpha ^{\prime },r_{\alpha
^{\prime }}}\mathbf{)}r_{0}^{2})=\mathcal{F}_{\mathbf{P}_{\alpha ,r_{\alpha
}}}(\mathbf{n\times (P}_{\beta ,r_{\beta }}\mathbf{-P}_{\alpha ,r_{\alpha }}%
\mathbf{)}r_{0}^{2})\varphi _{\mathbf{P}_{\beta ,r_{\beta }}}(\mathbf{0}).
\]
Similarly it also follows

\[
\varphi _{\mathbf{P}_{\alpha ^{\prime },r_{\alpha ^{\prime }}}}(-\mathbf{%
n\times (P}_{\beta ^{\prime },r_{\beta ^{\prime }}}\mathbf{-P}_{\alpha
^{\prime },r_{\alpha ^{\prime }}}\mathbf{)}r_{0}^{2})=\mathcal{F}_{\mathbf{P}%
_{\alpha ^{\prime },r_{\alpha ^{\prime }}}}(\mathbf{n\times (P}_{\beta
^{\prime },r_{\beta ^{\prime }}}\mathbf{-P}_{\alpha ^{\prime },r_{\alpha
^{\prime }}}\mathbf{)}r_{0}^{2})\varphi _{\mathbf{P}_{\beta ^{\prime
},r_{\beta ^{\prime }}}}(\mathbf{0}),
\]
which finally leads to the desired results

\begin{eqnarray}
&&\frac{\varphi _{\mathbf{P}_{\alpha ,r_{\alpha }}}^{*}(\mathbf{n\times (P}%
_{\beta ^{\prime },r_{\beta ^{\prime }}}\mathbf{-P}_{\alpha ^{\prime
},r_{\alpha ^{\prime }}}\mathbf{)}r_{0}^{2})\varphi _{\mathbf{P}_{\alpha
^{\prime },r_{\alpha ^{\prime }}}}^{*}(-\mathbf{n\times (P}_{\beta ^{\prime
},r_{\beta ^{\prime }}}\mathbf{-P}_{\alpha ^{\prime },r_{\alpha ^{\prime }}}%
\mathbf{)}r_{0}^{2})}{N_{\mathbf{P}_{\beta ^{\prime },r_{\beta ^{\prime
}}}}^{(2,3)}N_{\mathbf{P}_{\beta ,r_{\beta }}}^{(2,3)}}= \\
&&\frac{\mathcal{F}_{\mathbf{P}_{\alpha ,r_{\alpha }}}^{*}(\mathbf{n\times (P%
}_{\beta ,r_{\beta }}\mathbf{-P}_{\alpha ,r_{\alpha }}\mathbf{)}r_{0}^{2})%
\mathcal{F}_{\mathbf{P_{\alpha ^{\prime },r_{\alpha ^{\prime }}}}}^{*}(%
\mathbf{n\times (P}_{\beta ^{\prime },r_{\beta ^{\prime }}}\mathbf{-P}%
_{\alpha ^{\prime },r_{\alpha ^{\prime }}}\mathbf{)}r_{0}^{2})}{2\pi
r_{0}^{2}N_{\phi _{0}}^{2}}.  \nonumber
\end{eqnarray}
The expression for the matrix element (\ref{matrix1}) allows to directly
write formula (\ref{e2alter}) employed in Section 3 for evaluating the
correlation energy.


\begin{references}
\bibitem{stormer}  D.C. Tsui, H.L. Stormer and A.C. Gossard , {\it Phys.
Rev. Lett. }{\bf 48}, 1559 (1982).

\bibitem{laughlin}  R.B. Laughlin, {\it Phys. Rev. Lett. }{\bf 50}, 1395
(1983).

\bibitem{book}  O. Heinonen (Ed.), ''{\it Composite Fermions, an unified
view of the Quantum Hall effect regime'', }(World Scientific, Singapore,
1998).

\bibitem{clar1}  F. Claro, {\it Sol. State Commm. }{\bf 53}, 27 (1985).

\bibitem{clar2}  F. Claro, {\it Phys. Rev. }{\bf B35}, 7980 (1987).

\bibitem{yosh}  D. Yoshioka and P.A. Lee, {\it Phys. Rev. }{\bf B27},
4986(1983).

\bibitem{cabo1}  A. Cabo, {\it Phys. Lett.} {\bf A191}, 323 (1994).

\bibitem{cabclaper}  A. Cabo, F. Claro and A. Perez, Phys. Rev. {\bf B66},
035326-1(2002).

\bibitem{mikha}  S. A. Mikhailov, {\it Physica} {\bf B299}, 6 (2001).

\bibitem{clar3}  F. Claro, {\it Phys. Stat. Solidi} (b) {\bf 230}, 321 (2002)

\bibitem{he} S. He, {\it Computers in Physics} {\bf 11}, 194 (1997)

\bibitem{yosh1}  N. Shibata and D. Yoshioka, {\it Phys. Rev. Lett. }%
{\bf 86}, 5755 (2001). .

\bibitem{arov}  S. Kivelson, C. Kallin, D. P. Arovas and J. R. Schrieffer, {\it %
Phys. Rev. }{\bf B36}, 1620 (1987).

\bibitem{tesanovic}  B. I. Halperin, Z. Tesanovic and  F. Axel, {\it Phys. Rev. Lett. }%
{\bf 57}, 922 (1986).

bibitem{wannier}  G.H.Wannier, {\it Phys. Stat. Solidi }(b) {\bf 121}, 603
(1984).

\bibitem{ferrari}  R. Ferrari, {\it Phys. Rev. }{\bf B42}, 4598 (1990).

\bibitem{cabo0}  A. Cabo,{\it \ Phys. Lett. }{\bf A171}, 90 (1992).

\bibitem{cabo2}  A. Cabo,{\it \ Phys. Lett. }{\bf A211}, 297 (1996).

\bibitem{tao}  R. Tao, {\it J. Phys.}{\bf C19}, L619 (1986).

\bibitem{laugh}  R. B. Laughlin, in {\it ''The Quantum Hall Effect''},
R.E.Prange and S. Girvin, Eds, (Springer-Vaerlag, Berlin 1990) 231
\end{references}
\end{document}